\documentclass[aps,preprint,nofootinbib,floatfix]{revtex4}

\usepackage{amsmath,amssymb,epsfig,ulem,color,slashed}
\usepackage{xcolor}
\allowdisplaybreaks

\usepackage{booktabs}
\usepackage{hyperref}

\makeatother

\makeatletter 
\renewcommand{\fnum@figure}{\textbf{Figure~\thefigure}}
\makeatother

\allowdisplaybreaks  
\usepackage{tikz}
\usetikzlibrary{decorations.pathmorphing,patterns}
\def\ga   {\gamma}
\def\Ga   {\Gamma}

\def\la   {\lambda}

\def\sig   {\sigma}

\def\nn{\nonumber}
\def\lee { \left( }
\def\rii { \right) }

\def\De {\Delta}
\def\ka  {\kappa}

\def\toto {\leftrightarrow}
\def\Rp {R_{\psi\psi}}
\def\Rpb {R_{\bar{\psi}\bar{\psi}}}
\def\Rm {R_{\bar{\psi}\psi}}

\def\phip { \phi^\prime}

\definecolor{Red}{rgb}{1.,0.,0.}
\definecolor{Grn}{rgb}{0.,0.75,0.}
\definecolor{Blu}{rgb}{0.,0.,1.}

\newcommand{\AddrCP}{
CP$^{3}$-Origins, University of Southern Denmark, Campusvej 55
 DK-5230 Odense M, Denmark
}
\newcommand{\AddrDTU}{
Fakult\"at f\"ur Physik, TU Dortmund, D-44221 Dortmund, Germany
}

\begin{document} 

\title{Two-component asymmetric dark matter via bound states and freeze-in decay}

\author{Mathias Becker} \email{mathias.becker@tu-dortmund.de}\affiliation{\AddrDTU}
\author{Wei-Chih Huang} \email{huang@cp3.sdu.dk}\affiliation{\AddrCP}




\begin{abstract}
We propose a novel mechanism to realize two-component asymmetric dark matter of very different
mass scales through bound state formation and late freeze-in decay.
Assuming a particle-antiparticle asymmetry is initially shared by SM baryons and two dark matter components, we demonstrate that the existence of bound states formed by the heavy component can efficiently transfer the asymmetry
from the heavy to the light component via late decay. In this case, the energy densities of the two components can be comparable, and the correct relic density is reproduced.\\
{\footnotesize  \it Preprint: 
DO-TH 19/27 \; \; CP3-Origins-2019-43 DNRF90
}
\end{abstract}

\maketitle  

\section{Introduction}
\label{sec:introduction}

The identity of dark matter~(DM), an important missing piece in the standard model~(SM),
remains mysterious although the
astrophysical evidence of DM is well established. The DM relic
density is precisely known to be
$\Omega_{\text{DM}}=0.26$~\cite{Tanabashi:2018oca, Aghanim:2018eyx}, inferred from the
measurements of the power spectrum of the cosmic microwave background
radiation. Any realistic DM model has to reproduce this value. 
Furthermore, possibilities that DM consists of more than one species have been studied widely; 
for instance, multiple light species including neutrinos and axions~\cite{Hannestad:2003ye,Hannestad:2010yi} 
or in the context of supersymmetry involving axinos~\cite{Baer:2009ms,Bae:2013hma}.
Alternatively, all components can be Weakly Interacting Massive Particles~(WIMPs) whose stability is protected by a discrete symmetry, parity
or gauge symmetry; see, e.g., Refs.~\cite{Ma:2006km,Zurek:2008qg,Batell:2010bp,Fukuoka:2010kx,Belanger:2012vp,Aoki:2012ub,Ivanov:2012hc,Chialva:2012rq,Heeck:2012bz,Modak:2013jya,Aoki:2013gzs,Geng:2013nda,Kajiyama:2013rla,Bhattacharya:2013hva, Esch:2014jpa} and also Ref.~\cite{Belanger:2014vza} on
the classification of two-component DM models and relic density computation.

Models of two-component DM with very different masses, featuring boosted DM~\cite{Agashe:2014yua}, draw quite attention.
The heavy component, that can accumulate at the galactic center or be trapped around the center of the sun, 
annihilates into the highly relativistic light component, which in turn can enhance the DM-nucleon interaction rate
 at DM detectors~\cite{Agashe:2014yua,Berger:2014sqa, Kopp:2015bfa}.
To be more specific, a relativistic DM particle can yield a large momentum transfer in DM direct detection or up-scatter into
a heavier state via inelastic scattering~\cite{Giudice:2017zke} such that even DM of sub-GeV or below can be potentially probed in direct searches.
That is unlike the conventional scenarios with non-relativistic DM particles where the experimental sensitivity plummets for DM lighter than a few GeV.
On the other hand, the appealing idea of asymmetric dark matter~(ADM) has been proposed~\cite{Nussinov:1985xr}~(also Refs.~\cite{Zurek:2013wia, Petraki:2013wwa} for reviews) to link the DM relic density to the baryon asymmetry, that is also unaccountable in the SM framework. 
In this case, either DM particles or antiparticles remain in the universe due to a local or global asymmetry, analogous to the one that distinguishes
baryons from anti-baryons. In addition, generation mechanisms of baryon and DM asymmetries are often interwoven,
leading to roughly comparable amounts of asymmetry in the two sectors and
hence implying a DM mass of $\sim5$ GeV.

It is intriguing to meld these two ideas, i.e., two-component ADM of very different masses~($\sim$ GeV and $\gtrsim 100$ GeV respectively).
A grave repercussion will be an excessive relic density, if both the light and the heavy components have roughly the same 
amount of asymmetry as the SM baryons.
In other words, the resulting total DM relic abundance would overclose the universe as the heavy component
is by far too heavy to have a number density similar to that of the baryons. 
 The problem can be circumvented
if the  asymmetry generation mechanisms for two DM species are not  related or the amounts of asymmetry are controlled by independent parameters; for instance, they are generated by decays of two different heavy bosons or of the same heavy boson but with different couplings\,\footnote{It has been demonstrated that ADM can have a very different mass from GeV in the context of two-sector leptogenesis~\cite{Falkowski:2011xh,Falkowski:2017uya}, where the right-handed neutrinos decay both into the SM and DM particles with dissimilar amounts of asymmetry.
Models of ADM with a much heavier mass can also be realized in the context of bosonic technicolor~\cite{Frandsen:2019efk}.}.
In this situation, the number density of the heavy component can be arbitrarily small without exceeding the observed
DM density.

In this work, we explore an alternative scenario which employs bound state formation~(BSF) via a long-range interaction.
 The interaction arises when the underlying mediator is much lighter than the interacting particles, giving rise to the so-called Sommerfeld effect~(or Sommerfeld enhancement)~\cite{ANDP:ANDP19314030302,Sakharov:1948yq}.
 The enhancement increases the DM annihilation rate~\cite{Hisano:2002fk,Hisano:2003ec} and opens up new regions of the parameter space, 
previously not viable. Besides, induced BSF among DM particles and/or heavy states can also assist depleting the relic density~\cite{Ellis:2015vaa, Ellis:2015vna, Kim:2016zyy, Keung:2017kot}. 
Such the long-range interactions are considered in the context of co-annihilation~(with a slightly heavier but nearly degenerate partner~\cite{Binetruy:1983jf,
Griest:1990kh}) and also in supersymmetric models; see, e.g., Refs~\cite{Hisano:2002fk, Hisano:2003ec, Pospelov:2008jd, Hryczuk:2010zi, Hryczuk:2011tq, Beneke:2014hja, Beneke:2014gja, Harz:2014gaa, vonHarling:2014kha, Petraki:2015hla, Cirelli:2015bda, Pearce:2015zca,  
  An:2016kie, Petraki:2016cnz, Kouvaris:2016ltf, 
  ElHedri:2016onc, Liew:2016hqo, Asadi:2016ybp, An:2016gad, 
  Cirelli:2016rnw, Kim:2016kxt, Kim:2016zyy, Biondini:2017ufr, Baldes:2017gzu,Baldes:2017gzw, 
  Biondini:2018pwp, Biondini:2018xor, Biondini:2018ovz, Ellis:2018jyl, Geller:2018biy, 
  Harz:2018csl, Cirelli:2018iax, Bhattacharya:2018ooj, Schmiemann:2019czm}.
Recently, it has been pointed out that the Higgs boson can induce BSF as
long as it is much lighter than particles forming bound states~\cite{Harz:2017dlj, Harz:2019rro}.
Furthermore,  during the BSF process scattering with particles from the thermal bath
will significantly enhance the BSF cross-section in some cases~\cite{Binder:2019erp}. 

In our setup, there exist two DM sectors which separately contain the light and heavy component, $\chi$ and $\psi$.
We assume a particle-antiparticle symmetry was created by an unspecified mechanism at a early time and was~(roughly) equally
shared by $\chi$, $\psi$ and SM baryons.
The realization of two-component ADM of distinctive masses relies on
long-distance Yukawa interactions among particles of $\psi$ and $\bar{\psi}$.
The resulting $\psi$-$\bar{\psi}$ bound state facilitates the elimination of the symmetric component, while the
remaining asymmetric component consists of bare $\psi$ particles or bound states.
As we shall see below, the ratio of the number density of the bound states to that of the bare $\psi$ is controlled by the binding energy of the bound state.
The bound state will eventually {\it freeze-in}~\cite{McDonald:2001vt, Hall:2009bx, Bernal:2017kxu} decay to
a pair of $\chi$~(not a particle-antiparticle pair) via annihilations of the constituents
of the bound state. Therefore, the final densities of $\chi$ and $\psi$ will depend on the aforementioned ratio between the bare $\psi$ and the bound states before the decay.   
For sizable Yukawa couplings, the majority of the asymmetry of $\psi$ will be converted into that of $\chi$.
In the presence of  a disparity in the number densities, $n_\psi \ll n_\chi$, the corresponding
energy densities can still be of the same order, $\Omega_\chi \sim \Omega_\psi$, due to the distinct particle masses.

Note that this scenario is not the minimal setup to realize two-component ADM.
Instead of the freeze-in mechanism, one can, for instance, have standard freeze-out of annihilations of
$\psi$ into $\chi$ or SM fermions without involving bound states at all. The correct DM abundance can be reproduced by carefully tuning relevant coupling constants.  
We argue that long-range interactions~(bound states) themselves have rich and profound phenomenological implications. 
Moreover, there could exist situations where having  $\psi$ in thermal equilibrium at early times will erase the induced asymmetry.
To be more concrete, if annihilations of $\psi$ into SM fermions~($\psi\psi \leftrightarrow \bar{f}f$) and the asymmetry generation mechanism are present at the same time, there will be no initial asymmetry of $\psi$  and thus no two-component ADM.    

The paper is organized as follows. In Section~\ref{sec:boltz}, we briefly review the formalism of Boltzmann Equations which is used to obtain the time evolution
of the particle densities. Section~\ref{sec:model} will be devoted to detail a simple model and the sequence of the asymmetry shift among the different particle species.
Next, we show numerical results in Section~\ref{sec:result} and examine effects of several relevant parameters on the relic density. Four benchmark sets of the parameters are presented that can realize two-component ADM with comparable energy densities. Finally, we conclude in Section~\ref{sec:con}.
Computations of all relevant annihilation cross-sections as well as bound state formation and dissociation are collected in the Appendices.

\section{Boltzmann Equations}
\label{sec:boltz}

To begin, we shortly review the Boltzmann equations used to find the
time evolution of the various particle densities. More detailed
discussions can be found in Refs.~\cite{Griest:1990kh, Edsjo:1997bg,
  Giudice:2003jh}. Due to the expansion of the universe, a convenient
quantity to describe the particle number density is $Y \equiv
n/s_{\text{en}}$, the particle number density normalized to the
entropy density $s_{\text{en}}$, i.e., the number of particles per
comoving volume. The Boltzmann equation for the DM particle $\chi$
reads
\begin{align}
\label{eq:BoltzmannY}
	z H s_{\text{en}} \frac{d Y_{\chi}}{dz} = -\sum_{ \{a_i\}, \{f_j\} } [
          \chi a_1 \cdots a_n \leftrightarrow f_1 \cdots f_m ]\,, 
\end{align}
where $z = m_{\chi}/T$ and $H$ is the Hubble parameter, while  
\begin{align}
	[  \chi a_1 \cdots a_n \leftrightarrow f_1 \cdots f_m]=&  
	\frac{n_\chi n_{a_1} \cdots n_{a_n}}{n_\chi^{\rm eq} n_{a_1}^{\rm eq}\cdots n_{a_n}^{\rm eq}} 
	\gamma^\text{eq}(  \chi a_1 \cdots a_n \leftrightarrow f_1 \cdots f_m)                \nonumber\\  
	&- \frac{n_{f_1} \cdots n_{f_m}}{n_{f_1}^{\rm eq} \cdots n_{f_m}^{\rm eq}}
      \gamma^\text{eq}\left( f_1 \cdots f_m \leftrightarrow \chi a_1 \cdots a_n \right).
      \label{eq:ga_dif}
\end{align}
The symbol $\gamma^{\text{eq}}$ stands for the interaction rate in thermal equilibrium, defined
as
\begin{align}
	\gamma^{\text{eq}}(\chi a_1 \cdots a_n \to f_1 \cdots f_m) 
	&=      \int \frac{\mathrm{d}^3 p_{\chi}}{2 E_{\chi} (2\pi)^3} e^{-\frac{E_{\chi}}{T}} \times
	    \prod\limits_{a_i} \Big[ \int \frac{\mathrm{d}^3 p_{a_i}}{2 E_{a_i} (2\pi)^3} e^{-\frac{E_{a_i}}{T}} \Big] 
	 \nonumber\\
	 & \times \prod\limits_{f_j} \Big[ \int \frac{\mathrm{d}^3 p_{f_j}}{2 E_{f_j} (2\pi)^3}  \Big]
       \times (2\pi)^4\delta^4 \Big( p_{\chi} + \sum_{i=1}^n p_{a_i} - \sum_{j=1}^m p_{f_j} \Big) |M|^2\,,
\label{eq:ga_def}         
\end{align}
where $|M|^2$ is the squared amplitude summed over the initial and final
spins in the presence of fermions.
Note that in this work we always assume the absence of
tree-level CP violation, and hence $\gamma^{\text{eq}}(i j \cdots \to
k \chi \cdots)= \gamma^{\text{eq}}( k \chi \cdots \to i j
\cdots)$. For $2 \toto 2$ processes, the thermal rate can be succinctly expressed
as~\cite{Giudice:2003jh}
\begin{align}
 \ga^\text{eq} \lee a_1 a_2 \toto f_1 f_2 \rii = \frac{T}{64 \pi^4} \int^{\infty}_{s_{\text{min}}} ds \, \sqrt{s} \, 
 \hat{\sigma}(s) \, K_1\lee \frac{\sqrt{s}}{T} \rii ,
 \label{eq:ga_eq}
\end{align}  
where $s$ is the squared center-of-mass energy and $s_{\text{min}}=
\text{max}\left[ (m_{a_1}+m_{a_2})^2,(m_{f_1}+m_{f_2})^2\right]$.
The symbol $\hat{\sig}$ is the reduced cross-section defined as $\hat{\sig}
\equiv 2 s \, \la(1,m^2_{a_1}/s,m^2_{a_2}/s) \, \sig$, where
$\sig$ is the cross-section, summed over the initial and final spins, and  $\la$ is the
phase-space function: $\la[a,b,c] \equiv (a-b-c)^2 - 4 \, b \, c \,$. 
On the other hand, for the decay of the particle $a_1$, the thermal rate
becomes~\cite{Giudice:2003jh}
 \begin{align}
 \ga^{\text{eq}} \lee a_1 \toto f_1 f_2 \rii = n^{\text{eq}}_{a_1} \frac{K_1 \lee z \rii}{K_2 \lee z \rii} \Ga_{a_1},
 \label{eq:ther_decay}
 \end{align}
 where $z=m_{a_1}/T$, $\Ga_{a_1}$ is the decay width of $a_1$ at rest,
 and $K_i$ refers to the modified Bessel function of the $i$-th kind.

To account for the observed DM relic density,
$\Omega_{\text{DM}}=0.26$~\cite{Tanabashi:2018oca, Aghanim:2018eyx}, the requisite number
density in the comoving frame is
\begin{align}
Y_{\text{DM}} \lee z \to \infty \rii = \frac{4.32 \times 10^{-10}}{ \lee m_{\text{DM}}/ \text{GeV} \rii } \, ,
\label{eq:Y_DM}
\end{align}
where $m_{\text{DM}}$ is the DM mass.


\section{A simple model and the sequence of asymmetry transfer}\label{sec:model}

In this Section, we present a model which can accommodate two-component ADM $\chi$ and $\psi$ with very different masses
of $\sim$GeV and $\gtrsim 100$ GeV respectively, followed by detailed discussions on how the asymmetry is transferred among the different particle species. 

\subsection{Model}

There exist two dark sectors that contain vector-like fermion $\psi$ and $\chi$ respectively, both of which carry a charge of $+1$ under a global $U(1)'$ symmetry but are singlets
under the SM gauge groups. The $U(1)'$ charge ensures the DM stability
because all SM particles are neutral under the $U(1)'$. 
These two DM sectors are individually in thermal equilibrium with the SM sector via interactions of $\bar{\chi} \chi, \bar{\psi}{\psi} \leftrightarrow \bar{f} f$ that are assumed to 
be efficient enough  to deplete the symmetric components of
$\chi$ and $\psi$.
Additionally, there are two scalars $\phi$~(real) and $\phi'$~(complex). The particle $\phi$ is a pure singlet, and mediates the long-range interactions among $\psi$ and $\bar{\psi}$ particles, resulting in bound state formation~(BSF), $i + j \to [ij] + \phi$~($i$ and $j$ are referring to
$\psi$ and/or $\bar{\psi}$, and $[ij]$ is the bound state made of the fields $i$ and $j$) and the inverse process, bound state dissociation~(BSD). On the other hand, $\phi'$ has a $U(1)'$ charge of $-2$ and induces interactions that can shift asymmetry between $\chi$ and $\psi$\,\footnote{Instead of including $\phi'$,
one can assume feeble $\psi \psi \to \bar{f} f$ which breaks the $U(1)'$ symmetry and makes the bound state decay
at a later time, reducing $Y_\psi$ to achieve $\Omega_\psi \sim \Omega_\chi$.
In this case, the stability of $\psi$ is still protected by a residual $Z_2$ symmetry.
Notwithstanding, the freeze-in process of $\psi\psi \to \chi\chi$ under consideration can give rise to boosted $\chi$ which entails rich
phenomenological consequences.}.
The particle contents are summarized in Table~\ref{tab:PC}. The relevant Lagrangian reads 
\begin{align}
\mathcal{L} \supset & - y \, \phi \bar{\psi}  \psi - y' \, \phi \bar{f}  f 
- \kappa_\chi \, \phi' \overline{\chi^\text{c}} \chi - \kappa_\psi \, \phi' \overline{\psi^\text{c}}  \psi 
+ \frac{\bar{\chi} \ga^\mu \chi \bar{f} \ga_\mu f }{ \Lambda^2_\chi}  + \frac{\bar{\psi} \ga^\mu\psi \bar{f} \ga_\mu f }{ \Lambda^2_\psi}  \nn\\
& - m_\chi \bar{\chi} \chi - m_\psi \bar{\psi} \psi - \frac{1}{2} m^2_\phi \phi^2 - m^2_{\phi'} \phi'^* \phi' \, 
\label{eq:rel_lan}
\end{align} 
where the superscript $c$ denotes charge conjugate that explicitly
indicates that the $\phi'$-Yukawa interactions engender the asymmetry transfer,  $\chi \chi~(\bar{\chi}\bar{\chi}) \leftrightarrow \psi \psi~(\bar{\psi}\bar{\psi})$.
The two four-fermion effective operators characterize interactions between DM and SM fermions~($f$), mediated by an unspecified heavy gauge boson.
These interactions not only keep both $\chi$ and $\psi$ in the thermal bath but also eliminate the symmetric components of $\chi$ and $\psi$\,\footnote{The heavy component $\psi$
has an additional annihilation channel $\bar{\psi} \psi \to \phi\phi$.} when $T \lesssim m_\chi \, , m_\psi$ which leads to ADM. 
For $m_\chi \sim$ GeV and $m_\psi \sim$ TeV, one requires $\Lambda_\chi \lesssim 350 \, \mathrm{GeV}$ and $\Lambda_\psi \lesssim 6.5 \, \mathrm{TeV}$.
to annihilate away the symmetric components. To avoid or mitigate potential experimental constraints
on the underlying gauge boson, one can simply assume the gauge boson only couples to the third-generation 
SM fermions. The identity of the gauge boson and the details of the interactions are not relevant to our discussions
as long as the resulting cut-off scales $\Lambda_\chi$ and $\Lambda_\psi$ are small enough to keep the interactions
in equilibrium.

The Yukawa coupling $y'$ leads to decays of $\phi$ into SM fermions\,\footnote{The coupling can arise, e.g., via the mixing
between the SM Higgs boson and $\phi$, leading to $y' \approx  y^{\text{SM}}_f  \sin\theta$, where $\theta$ is the mixing angle and $y^{\text{SM}}_f$ is the SM Yukawa coupling.
To ensure $\phi$ is in the thermal bath for $T \lesssim m_\phi \sim$ GeV~(scale of interest in this work), we have $y' \gtrsim 10^{-9}$,
indicating $\sin\theta \gtrsim 10^{-7}$, depending on the final state fermions.
Such a small mixing angle is well below the current experimental sensitivity.}
if kinematically allowed as well as keeping $\phi$ in thermal equilibrium
until $T$ drops well below $m_\phi$.
\begin{table}[htp!]
	\centering
	\begin{tabular}{|c|c|c|c|c|}
		\hline
		& $\chi$ & $\psi$ & $\phi$ & $\phi'$ \\
		\hline
		Mass & $\mathcal{O} \left( \mathrm{GeV} \right) $ & $\mathcal{O} \left( \gtrsim 100~\mathrm{GeV} \right) $ & $\lesssim \mathrm{GeV}$ & $> \mathrm{TeV}$ \\
		\hline
		$U\left( 1 \right)_D$ & $+1$ & $+1$ & $0$ & $-2$ \\
		\hline
	\end{tabular}
	\caption{The particle contents in the dark sectors where all particles are singlets under the SM gauge groups. See the text for details.}
	\label{tab:PC}
\end{table}
As mentioned above, the bound states form in the $\psi$ sector due to the Yukawa interaction of $\phi$.
Since the interaction is always attractive among particles and antiparticles,
there exist three types of the bound states: $\Rp~(\sim[\psi\psi])$, $\Rpb~(\sim [\bar{\psi} \bar{\psi}] )$ and $\Rm ~ ( \sim [\psi \bar{\psi}])$.
In the limit of $m_\phi \ll y^2 m_\psi/ \left( 8 \pi \right)$~(inverse of the Bohr radius), the Yukawa potential can be well approximated by a Coulomb potential.
This approximation
greatly simplifies the calculations on the cross-sections of BSF and BSD,
that are summarized in Appendix~\ref{app:Bound_F_D}.
In this work, we consider only the ground state with a binding energy:
\begin{align}
\label{eq:binding}
 E_B = -   \frac{y^4}{64 \, \pi^2} m_\psi + \frac{y^2}{4 \, \pi}  m_\phi.
 \end{align}
The bound state mass is $m_R = 2 m_\psi + E_B$,
with $m_{\Rp}$$=$$m_{\Rpb}$$=$$m_{\Rm} \equiv m_R$.

The bound states themselves will not be stable as $\Rm$ will quickly decay either into a pair of $\phi$ induced by the $\phi$-Yukawa interaction or 
into SM fermions via the annihilations of $\psi$ and $\bar{\psi}$,
while $\Rp$~($\Rpb$) will eventually decay into a pair of $\chi$~($\bar{\chi}$)
through the feeble interactions
$\chi \chi \leftrightarrow \psi \psi$~($\bar{\chi}\bar{\chi} \leftrightarrow \bar{\psi}\bar{\psi}$).
In light of the asymmetry of $\psi$, only $\Rp$~($\Rpb$) exists for $T \ll m_\psi$ in case $Y_\psi >  Y_{\bar{\psi}}~(Y_\psi <  Y_{\bar{\psi}})$ initially. 
All relevant  cross-sections and decay rates are given in Appendix~\ref{app:reduced-X}.

In the following numerical analysis, we solve the Boltzmann equations for the species
$\phi$, $\psi$, $\bar{\psi}$, $\Rp$, $\Rpb$ and $\Rm$ by including the interactions of
BSF, BSD, $\bar{\psi} \psi \leftrightarrow \bar{f} f$,
$\bar{\psi} \psi \leftrightarrow \phi\phi$, $\Rm \leftrightarrow \phi\phi$, $\Rp \leftrightarrow \chi\chi$~($\Rpb \leftrightarrow \bar{\chi}\bar{\chi}$) and $\phi \leftrightarrow \bar{f} f$.

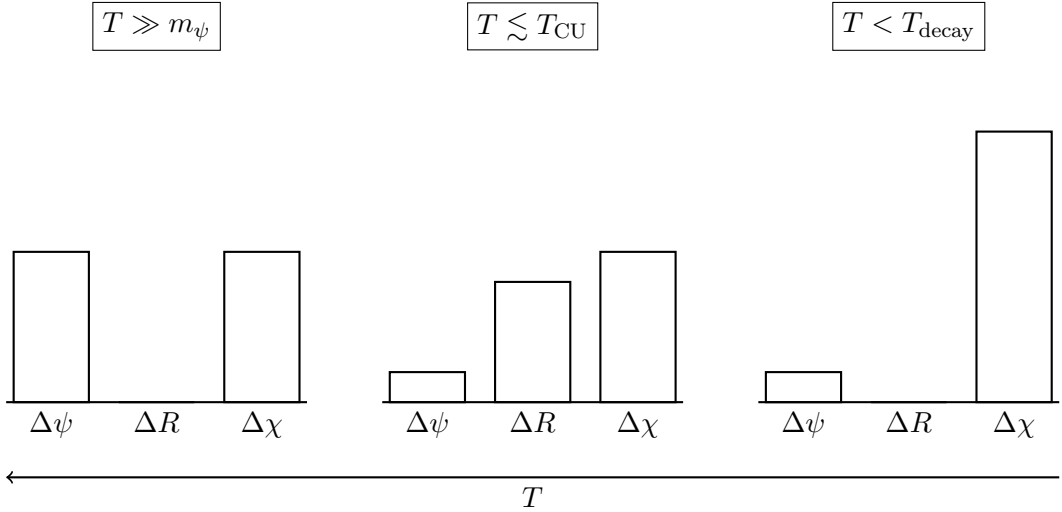
\begin{figure}
	\begin{center}
		\begin{tikzpicture}
		\node[draw] at (0,5) {$T \gg m_\psi$};
		\draw[thick] (-2,0) -- (2,0);
		\draw[thick] (-1.9,0) rectangle (-0.9,2);
		\node[below] at (-1.4,0){$\Delta \psi$};
		\draw[thick] (-0.5,0) rectangle (0.5,0);
		\node[below] at (0,0){$\Delta R$};
		\draw[thick] (0.9,0) rectangle (1.9,2);
		\node[below] at (1.4,0){$\Delta \chi$};
		
		\node[draw] at (5,5) {$T \lesssim T_\text{CU}$};
		\draw[thick] (3,0) -- (7,0);
		\draw[thick] (3.1,0) rectangle (4.1,0.4);
		\node[below] at (3.6,0){$\Delta \psi$};
		\draw[thick] (4.5,0) rectangle (5.5,1.6);
		\node[below] at (5,0){$\Delta R$};
		\draw[thick] (5.9,0) rectangle (6.9,2);
		\node[below] at (6.4,0){$\Delta \chi$};
		
		\node[draw] at (10,5) {$T < T_\text{decay}$};
		\draw[thick] (8,0) -- (12,0);
		\draw[thick] (8.1,0) rectangle (9.1,0.4);
		\node[below] at (8.6,0){$\Delta \psi$};
		\draw[thick] (9.5,0) rectangle (10.5,0);
		\node[below] at (10,0){$\Delta R$};
		\draw[thick] (10.9,0) rectangle (11.9,3.6);
		\node[below] at (11.4,0){$\Delta \chi$};
		
		\draw [<-,thick] (-2, -1) -- (12, -1) ;
		\node[below] at (5,-1){$T$};
		\end{tikzpicture}
	\end{center}
	\caption{Illustration of the asymmetry transfer between the heavy ($\psi$) and light ($\chi$) components with the help of the bound states $R$. At a high temperature, an asymmetry is generated and shared by both the dark and SM sectors. Below the temperature $T_\text{CU}$~(catch-up temperature at which $Y_R = Y_\psi/2$), more than half of the $\psi$ asymmetry has been stored in the bound states. The bound states later decay into $\chi$, thereby transferring the majority of the asymmetry from $\psi$ into $\chi$, leading to $Y_\chi \gg Y_\psi$ but with $\Omega_\chi \approx \Omega_\psi$ as $m_\psi \gg m_\chi$.}
	\label{fig:AsymmetryTransfer}
\end{figure}

\subsection{Asymmetry transfer}

\begin{figure}[htbp!]
\centering
\includegraphics[width=0.6\textwidth]{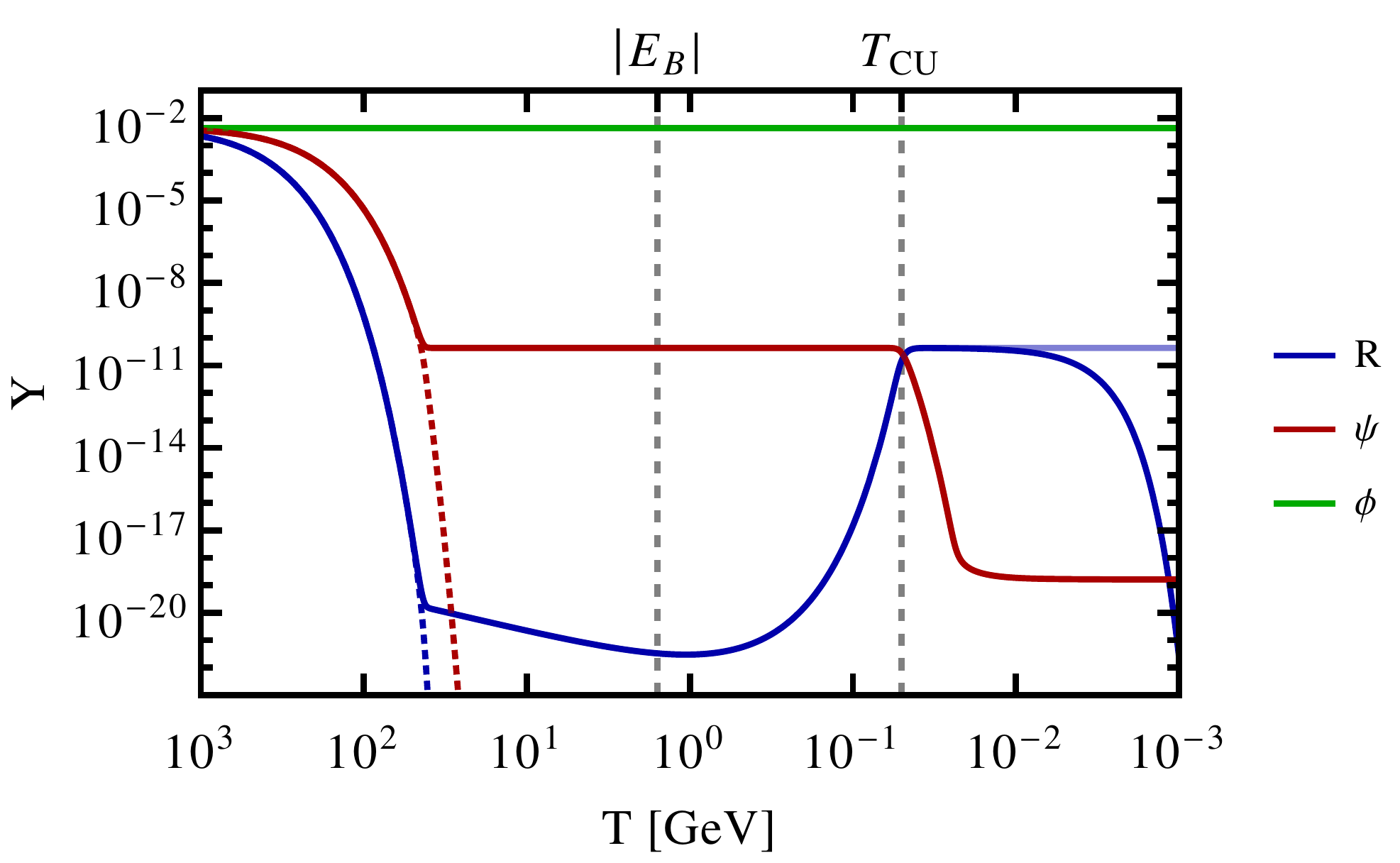}
\caption{The time evolution of the densities for particles in the dark sectors,
where  $(m_\psi, m_{\phi'}, \Lambda_\psi) = (1, 10, 10)$ TeV with massless $\phi$ and
 $(y, \ka_\chi, \ka_\psi) = (1, 10^{-4}, 10^{-4})$ are assumed. The solid green~(red, blue) line represents
 $Y_{\phi}$~($Y_{\psi}$,  $Y_{\Rp}$).  The dashed red~(blue) line refers to
 $Y_{\bar{\psi}}$~($Y_{\Rpb}$ and $Y_{\Rm}$), demonstrating that the symmetric components are annihilated away.
 The horizontal light blue line at small $T$ stands for the case of stable $\Rp$ without decaying into $\chi$, while the final
 density of $\psi$~(red line) is unaffected by the decay as BSF already stops before the decay. 
 }
	\label{fig:exam_y}
\end{figure}

We here elaborate in detail on how the initial asymmetry is transferred between the $\chi$ and $\psi$ sector as the universe cools down. 
 The sequence of the asymmetry transfer via BSF and BSD is pictorially illustrated in Fig.~\ref{fig:AsymmetryTransfer}.
 The time evolution of the densities of the relevant
 species are shown in Fig.~\ref{fig:exam_y}, in which for demonstration we choose $(m_\psi, m_{\phi'}, \Lambda_\psi) = (1, 10, 10)$ TeV with massless $\phi$ and $f$,
and $(y, \ka_\chi, \ka_\psi) = (1, 10^{-4}, 10^{-4})$, implying a binding energy of $\vert E_B \vert = 1.58$ GeV.
The solid green~(red, blue) line corresponds to $Y_\phi$~($Y_\psi$, $Y_{\Rp}$) while the dashed red~(blue) line represents $Y_{\bar{\psi}}$~($Y_{\Rpb}$ and $Y_{\Rm}$\footnote{The lines corresponding to $Y_{\Rpb}$ and $Y_{\Rm}$ are on top of each other.}).
The vertical dashed grey lines mark the absolute value of the binding energy and the catch-up temperature, defined below.
For $T \lesssim 10^{-2}$ GeV, the light~(dark) blue line refer to the case of stable~(decaying) $\Rp$.

\begin{itemize}
	\item
	At $T \gg m_{\psi}$, $\chi$ and $\psi$ are individually in thermal equilibrium with the SM.
	An unspecified mechanism is presumed for generating asymmetries in all $\chi$, $\psi$ and
	the SM baryons~(e.g., out-of-equilibrium decays of heavy gauge or Higgs
	bosons~\cite{Yoshimura:1978ex,Toussaint:1978br,Weinberg:1979bt,Barr:1979ye}).
	For simplicity, we assume that the total initial asymmetry of the three sectors adds up to zero:
	\begin{align}
	\Delta Y_B + \Delta Y^i_{\psi} + \Delta Y^i_{\chi} = 0 \, ,
	\label{eq:ini_as}
	\end{align} 
	in which the superscript $i$ refers to the initial values and $\Delta Y_f \equiv Y_{f} - Y_{\bar{f}}$.
	Furthermore, it is assumed that the generated baryon asymmetry accounts for the observed value, i.e. $\Delta Y_B  = \left( 8.6 \pm 0.7 \right) \times 10^{-11}$ \cite{Aghanim:2018eyx} and remains constant
	once being created\,\footnote{In fact, our conclusions do not rely on these assumptions.}.
	In this work, we set $\Delta Y^i_{\psi} \sim \Delta Y^i_{\chi} >0$, namely there are more $\psi~(\chi)$ than
	$\bar{\psi}~(\bar{\chi})$.

	\item 
	Depending on the binding energy and the mass of $\phi$, BSF and BSD are virtually efficient for a large part of the time period of interest.
	From Eq.~\eqref{eq:ga_dif}, it implies
		\begin{align}
	\frac{n^2_{\psi}} { \lee n^{\text{eq}}_{\psi} \rii^2}
	\approx  \frac{n_{\Rp}} { n^{\text{eq}}_{\Rp} } \frac{n_{\phi}} { n^{\text{eq}}_{\phi} }
	\, .
	\label{eq:eqcond}
	\end{align}
	 That in turn indicates $R$ also follows its equilibrium density when $T \gtrsim m_\psi$
	 because both $\psi$ and $\phi$ are in thermal equilibrium.
	
	\item For $ \vert E_B \vert  \lesssim  T \lesssim m_\psi $, the annihilations of $\bar{\psi}$ and $\psi$ into $\phi$ and SM fermions are kinematically more favorable than the reverse reactions and thus the number density of $\psi$ experiences a Boltzmann suppression.
	At a certain point, the equilibrium number density of $\psi$ becomes smaller than the asymmetry stored in $\psi$. It indicates that the symmetric component has been mostly obliterated and what remains is the asymmetric component -- $\psi$ particles. 
        The depletion of the symmetric component of the bound states also takes place roughly at the same time as $\psi$
        since the number densities are connected by Eq.~\eqref{eq:eqcond}.
        In our example, it happens around $T = 41 \, \mathrm{GeV}$ but with $Y_{\Rp} \ll Y_{\psi}$  because the former suffers a double Boltzmann suppression due to  $\exp ( - m_{R}/T) \approx \exp ( - 2m_{\psi }/T) \ll \exp ( - m_{\psi }/T)$. Alternatively, the relative suppression can be understood by inspecting Eq.~\eqref{eq:eqcond}: $n_R  \sim (m_\psi T)^{-3/2} (n_\psi)^2 \ll n_\psi $, given $n_{\phi}=n^{\text{eq}}_{\phi}$.

	\item At $T \lesssim |E_B|+ m_\phi$, the density $Y_{\Rp}$ increases sharply, catching up with $Y_\psi$,
	as clearly shown in Fig.~\ref{fig:exam_y}. One can apprehend the catch-up behavior via
	the interplay between BSF and the conservation of the total asymmetry as follows. 
	While the process of $\psi \psi \leftrightarrow \Rp \, \phi$ can change the number densities of $\Rp$ and $\psi$ individually, the total asymmetry stays constant before the decay of $\Rp$: 
	\begin{align}
	\Delta Y^i_{\psi} = Y_\psi + 2 Y_R \, .
	\label{eq:psiA}
	\end{align}
	Furthermore, as long as the interaction  $\phi \bar{f}f$ in Eq.~\eqref{eq:rel_lan} is faster than the expansion rate of the universe, $\phi$ is in equilibrium, implying $n_\phi = n^{\text{eq}}_\phi$. 
	As a result, combining Eq.~\eqref{eq:eqcond} and \eqref{eq:psiA},
	 one obtains an analytic expression for the number density of $\Rp$, provided that BSF and BSD are effective:
	\begin{align}
	Y_R = \frac{\Delta Y^i_{\psi}}{2} + \mathcal{R} \left( 1 - \sqrt{1 + \frac{\Delta Y^i_{\psi}}{\mathcal{R} }} \right) \, , 
	\label{eq:YR}
	\end{align}
	with
	\begin{align}
	\mathcal{R} = \frac{ \left( n_\psi^\text{eq} \right)^2}{8 n_R^\text{eq} s_\text{en}} \, .
	\end{align}
	The equilibrium number densities of the non-relativistic $\psi$ and $\Rp$ scale as
	$n^{\text{eq}} \sim \left( m T \right)^\frac{3}{2} \exp \left( - \frac{m}{T} \right)$.
	That implies $\mathcal{R} 
	\sim (m_\psi/T)^{-\frac{3}{2}} \exp \left( - \frac{\vert E_B \vert }{T} \right) \rightarrow 0$ 
	in the limit of $T \rightarrow 0$, and consequently $Y_R \overset{T \rightarrow 0}{=} \frac{\Delta Y^i_{\psi}}{2}$.
In other words, BSF is favored over BSD and most of the asymmetry would be transferred to the bound states. The underlying reason is that  more and more $\phi$ particles no longer have sufficient energy to overcome the binding energy~(a requirement to break the bound state),
when the temperature falls below $\vert E_B\vert$.  With a larger Yukawa coupling~(larger $\vert E_B \vert$),
	 more $\psi$ particles will be converted into $\Rp$, leading to more asymmetry being stored in $\chi$
	 after $\Rp$ decays.
	 That is why the existence of the bound states naturally allows for two-component ADM of very different mass scales but with comparable energy densities.
	 In fact, the situation here is  very similar to the one during recombination at which electrons and protons first became bound to form neutral hydrogen atoms.
	In case $\phi$ is massive,  its number density also has an exponential suppression at  $T< m_\phi$
	 such that in the end there are not enough $\phi$ particles to fragment the bound states, rendering BSD ineffective.

\item	We define the catch-up temperature $T_\text{CU}$ as the temperature when the asymmetry is equally shared by 
$\Rp$ and free $\psi$, i.e., $Y_\psi/2 = Y_{\Rp}$ at $T=T_\text{CU}$.
By setting $Y_R = \Delta Y^i_{\psi} / 4$ in Eq.~\eqref{eq:YR}, the value of $T_\text{CU}$ can be numerically calculated.
In our exemplary case shown in Fig.~\ref{fig:exam_y}, we have $T_{\text{CU}} = 0.05$ GeV.
With $m_\phi=0$, an empirical expression of $T_{\text{CU}}$  for $0.1 \lesssim y \lesssim 5$ can be found as
\begin{align}
T_\text{CU} \approx 0.03 \, |E_B| \, y^{1/5} \, ,
\label{eq:TCU}
\end{align}
with an accuracy above $90 \, \%$.

\item As more and more $\psi$ particles are being converted, it becomes harder and harder for them to find each other to form the bound state. 
Eventually below a certain temperature, denoted by $T_\text{D}$, the precise value of which is determined by
the parameters $m_\phi$ and $y$, BSF ceases to work, similar to the freeze-out of thermal DM.
In our example, $T_D$ is around $12 \, \mathrm{MeV}$ below which $Y_\psi$ stops decreasing and levels off
as shown in Fig.~\ref{fig:exam_y}.

The asymmetry stored in the bound states after the asymmetry transfer is given by $Y_{\Rp} \left(T_\text{D} \right)$, while the final asymmetry 
stored in $\psi$, represented by $\De Y^f_\psi$, is simply $Y_\psi(T_\text{D})$.
After the decay of $\Rp$, the final $\chi$ asymmetry is: 
	\begin{align}
	\Delta Y^f_{\chi}  = -\Delta Y_B - \Delta Y^f_{\psi}  \, ,
	\end{align}
	where we have used Eq.~\eqref{eq:ini_as} and
	\begin{align}
	\Delta Y^i_{\chi} + \Delta Y^i_{\psi} = \Delta Y^f_{\chi} + \Delta Y^f_{\psi} \, .
	\end{align}
	As a result, the ratio of the total energy density  of DM to that of the SM baryons reads
	\begin{align}
	\frac{\Omega_\text{DM}}{\Omega_B} =\left| \frac{\Delta Y^f_{\psi}}{\Delta Y_B} \right| \frac{m_\psi - m_\chi}{m_B} + \frac{m_\chi}{m_B} \, ,
	\end{align}
	and in order to reproduce the observed relic density one has 
	\begin{align}
	\left| \frac{\Delta Y^f_{\psi}}{\Delta Y_B} \right| = \frac{m_B}{m_\psi - m_\chi} \left( \frac{\Omega_\text{DM}}{\Omega_B} - \frac{m_\chi}{m_B} \right) \overset{m_\psi \gg m_\chi}{=} \frac{m_\chi}{m_\psi} \left[ \frac{m_B}{m_\chi} \frac{\Omega_\text{DM}}{\Omega_B} - 1  \right] \, . 
	\end{align}
	If we further require that the energy densities of $\psi$ and $\chi$ are comparable~($m_\psi \Delta Y_{\psi,f} \sim m_\chi \Delta Y_{\chi,f}$),
	the  mass of $\chi$ can be inferred
	\begin{align}
	m_\chi \sim \frac{m_\psi m_B }{2 m_\psi - m_B \frac{\Omega_\text{DM}}{\Omega_B}} \frac{\Omega_\text{DM}}{\Omega_B} \overset{m_\psi \gg m_B}{=} m_B  \frac{\Omega_\text{DM}}{2 \, \Omega_B} = 2.66 \, \mathrm{GeV} \, ,
	\end{align}
	that indicates
	\begin{align}
	\left| \frac{\Delta Y^f_{\psi}}{\Delta Y_B} \right|  \sim \frac{m_\chi}{m_\psi} ,
	\label{eq:fin_psi_B}
	\end{align}
given $\Omega_\text{DM} = 5.4 \, \Omega_{B} $ and $m_B \approx 1$ GeV.
Therefore, the final density of $\psi$~(solid red line)
in Fig.~\ref{fig:exam_y} is too small to have any impact on the relic density. 
\end{itemize} 

To sum up, when the temperature falls below $m_\psi$, the symmetric components of $\psi$ and $R$
will be destroyed by the processes of $\bar{\psi} \psi \to \phi\phi \, , \bar{f}f$ and $\Rm \to \phi\phi$,
and only the asymmetric component, comprising $\psi$ and $\Rp$, is left with $Y_\psi \gg Y_{\Rp}$ in light of the Boltzmann suppression.
At $T \lesssim \vert E_B \vert$, $Y_{\Rp}$ begins to catch up with $Y_\psi$ due to
the lack of energetic $\phi$ to dissociate the bound state, i.e., BSF being kinematically preferred over BSD.
With the continuous decrease in $Y_\psi$, BSF will ultimately become ineffective
as the BSF rate is proportional to $Y^2_\psi$,
Afterwards, the density $Y_\psi$ stays constant while the bound state decays into a pair of $\chi$. 
The Yukawa coupling $y$ determines when BSF stops and therefore the final value of $Y_\psi$.
In the following, we will discuss the sufficient conditions for realizing two-component ADM with
comparable energy densities but very different mass scales between the two DM components.
Throughout our analyses, we always assume $m_\chi=2.66$ GeV unless otherwise stated.

\section{Numerical results}\label{sec:result}
In this Section we present numerical solutions of the coupled Boltzmann equations involving
(anti-)particles of $\psi$ and $R$ as well as $\phi$.
We investigate how the Yukawa coupling $y$, the mediator mass $m_\phi$ and the decay width $\Gamma_{\Rp}$ individually influence
the final amounts of asymmetry stored in $\psi$ and $\Rp$~($\chi$). Finally, we present four benchmark scenarios with different $m_\psi$ but fixed $m_\chi$, where the correct relic density is reproduced with $\Omega_\chi \approx \Omega_\psi$.

\subsection{Effect of $y$ values} \label{sec:bsy}

\begin{figure}[htbp!]
\centering
	\includegraphics[width=0.7\textwidth]{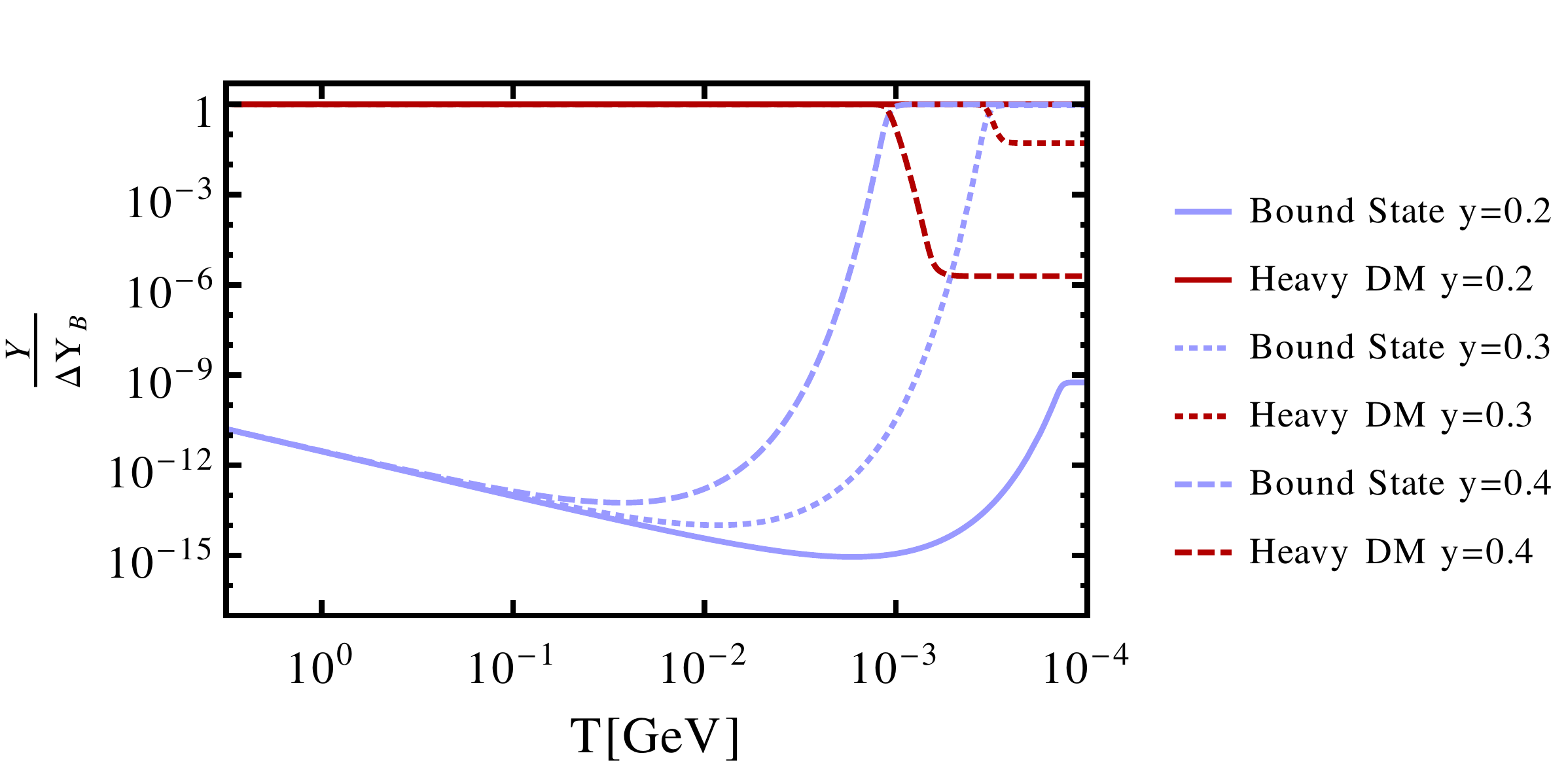}
	\caption{The results of the Boltzmann equations for different Yukawa couplings $y=(0.2,0.3,0.4)$, given
	$m_\chi=2.66 \, \mathrm{GeV}$, $m_\psi=1 \, \mathrm{TeV}$ and $\Gamma_{\Rp}=0$ with massless $\phi$.
	The red~(blue) lines represent the number
	density of $\psi$~($\Rp$) normalized to the baryon density.
	The different line styles correspond to the different values of $y$.
	Comparable densities $\Omega_\chi \approx \Omega_\psi$ can be accommodated  by taking $ y \approx 0.33$.
	The initial condition of $\Delta Y^i_{\psi} = \Delta Y_B$ at large $T$ is presumed. }
\label{fig:yPlots}
\end{figure}

In Fig.~\ref{fig:yPlots} we show the impact of different Yukawa couplings $y = (0.2, 0.3, 0.4 )$ on the final
number densities, assuming massless $\phi$, 
TeV $\psi$ and a stable bound state with an initial condition of $\Delta Y^i_{\psi} = \Delta Y_B$ at high $T$.
Clearly, the final  abundance of $\psi$ decreases as $y$ increases.
Since the BSF cross-section scales as $y^{12}$, a larger Yukawa coupling corresponds to a much larger BSF rate
and thus more $\psi$ form bound states, implying a smaller final density of $\psi$. 
On the other hand, from Eq.~\eqref{eq:TCU}
the catch-up temperature is proportional to $\vert E_B\vert y^{1/5} \sim y^{21/5}$ and hence
larger $y$ indicates an earlier catch-up as shown in Fig.~\ref{fig:yPlots}.
In the case of $y=0.2$, the BSF processes even cease to work prior to the catch-up.
As a consequence, to reproduce the correct relic density, most of the $\psi$ asymmetry has to be transferred into the bound
state, setting a lower bound on the value of $y$.

Given $m_\chi=2.66$ GeV and $m_\psi = 1$ TeV,
the value of $y  \sim 0.33$ is required to reproduce a correct value of $\Delta Y^f_\psi$ that fulfills Eq.~\eqref{eq:fin_psi_B}
and in turn attains $\Omega_\psi \sim \Omega_\chi$.
The corresponding plummet of $Y_\psi$~(surge on $Y_{\Rp}$) takes place between those of $y=0.3$ and $y=0.4$, i.e,
around $T \lesssim 1$ MeV. 
The bound states will eventually decay, creating a population of highly energetic $\chi$ particles
below the scale of Big Bang nucleosynthesis~(BBN).
Moreover, if the presumed asymmetry generation mechanism instead creates more $\bar{\chi}$ than $\chi$, then annihilations of induced $\chi$ with pre-existing $\bar{\chi}$ into the SM fermions will also inject sizable entropy into the thermal bath and thus the model will be constrained by BBN measurements. See, for instance, Refs.~\cite{Boehm:2012gr, Nollett:2014lwa, Hufnagel:2017dgo}.  

Since most of the bound states are produced around $T = T_{\text{CU}}$, in order not to perturb BBN  we have to raise
$T_{\text{CU}}$ above the scale of MeV such that the subsequent decay of the bound state can happen before the onset of BBN.
Naively thinking, one may enlarge $y$, which in return leads to an earlier catch-up and also puts an end to BSF above the BBN scale.
The increase on $y$, nonetheless, will also decrease significantly the final density $Y_\psi$ -- 
there is a difference of more than four orders of magnitude in $Y^f_\psi$ between the cases of $y=0.3$ and $y=0.4$,
resulting in $\left| \Delta Y^f_{\psi} / \Delta Y_B \right| \ll m_\chi/m_\psi$ and hence $\Omega_\psi \ll \Omega_\chi$.
One possible solution is to make $\phi$ massive in combination with a large value of $y$ as discussed below.

\subsection{Effect of a non-zero mediator mass} \label{sec:mediator}

\begin{figure}[htbp!]
\centering
	\includegraphics[width=0.7\textwidth]{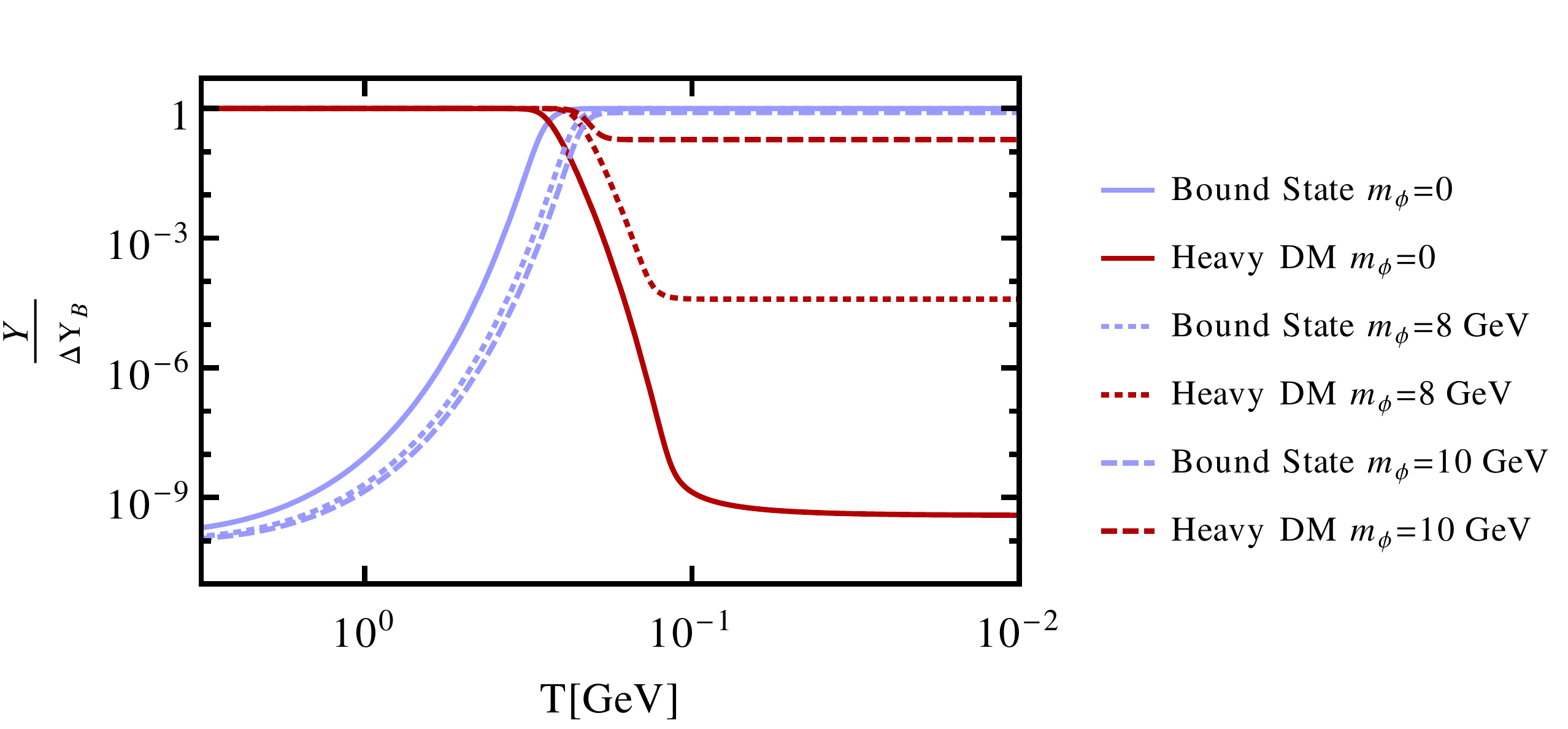}
	\caption{The solutions of the Boltzmann equations for $m_\psi=1 \, \mathrm{TeV}$, $y=1.5$ and the different values of $m_\phi= \left(0, 8, 10 \right) \, \mathrm{GeV}$. The binding energy, that partially depends on $m_\phi$ as shown in Eq.~\eqref{eq:binding}, is around 8 GeV.
	The red~(blue) lines represent the number density of $\psi$~($\Rp$) normalized to the baryon asymmetry. The different styles of the lines correspond to the different mediator masses.
	The density $Y^f_\psi$ increases with larger $m_\psi$, and the mass dependence is very striking.}
	\label{fig:mfPlot}
\end{figure}
 
The final densities of $\psi$ and $\Rp$ depend on the value of decoupling temperature $T_{\text{D}}$ below which BSF stops.
In case of a massless $\phi$, the BSF process $\psi \psi \to \Rp \phi$ is always kinematically allowed~($2 \, m_\psi > m_{R}$)  but it becomes ineffective when $Y_\psi$ is diminutive,
as explained above. To obtain a large value of $Y^f_\psi$, it is necessary to halt BSF earlier.
With $m_\phi > \vert E_B\vert$, in addition to the Boltzmann suppression from the density $n_\psi$,
BSF will also have a kinematical suppression by virtue of $2 \, m_\psi < m_{R} + m_\phi$ when $T < m_\phi$,
leading to a higher $T_{\text{D}}$ and hence a larger $Y^f_\psi$.
 In Fig.~\ref{fig:mfPlot}, it is clear that a larger mass of $\phi$ gives rise to a larger final density of $\psi$ where we fix $y=1.5$ and $m_\psi =1$ TeV
 with the same initial condition $\Delta Y^i_{\psi} = \Delta Y_B$ as above.
 The dependence of the final density $Y^f_\psi$ on the mass $m_\phi$ is quite remarkable -- 
 increasing $m_\phi$ from 8 to 10 GeV is accompanied by a factor of nearly $10^4$ on the final density.

On the other hand, the temperature $T_{\text{CU}}$ becomes lower in the presence of massive $\phi$ as shown in Fig.~\ref{fig:mfPlot}.
That can be understood by noticing that $Y_{\Rp}$ begins to rise when $\phi$ no longer has enough energy to break apart the bound states.
Due to the fact that  the mass of $\phi$ itself as energy can also be used to destroy the bound states, involving a
massive $\phi$ will postpone the catch-up and thus lower $T_{\text{CU}}$.
As a result, one would need a large value of $y$ together with a nonzero mass of $\phi~(\gtrsim \vert E_B\vert)$ to increase both $T_{\text{CU}}$ and $T_{\text{D}}$, ensuring the majority of the bound states decay before BBN
while attaining a sizable final density of $\psi$.

\subsection{Effect of a non-zero Decay Width}
\label{sec:DW}

 \begin{figure}[htbp!]
\centering
 	\includegraphics[width=0.7\textwidth]{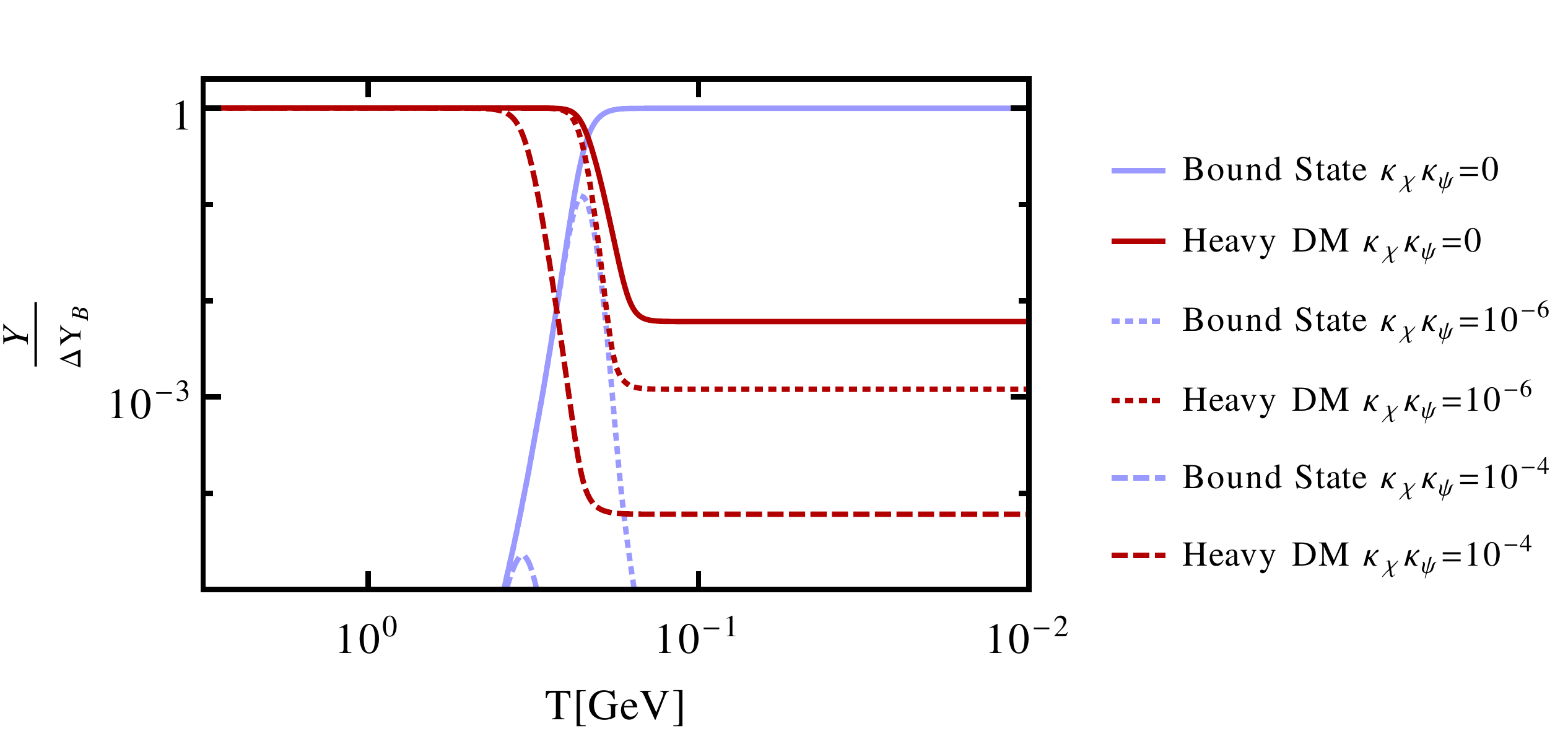}
 	\caption{The results of the Boltzmann equations, given $m_\psi=1 \, \mathrm{TeV}$, $m_\phi= 9 \, \mathrm{GeV}$, $y=1.5$ and different values of $\kappa_\chi \kappa_\psi = \left( 0, 10^{-6}, 10^{-4} \right)$, that correspond to different decay widths $\Gamma_{\Rp}$ as $\Gamma_{\Rp} \sim \left( \kappa_\psi \kappa_\chi \right)^2$. The red~(blue) lines represent the number density of $\psi$~($\Rp$) normalized to the baryon density.}
 	\label{fig:RPlot}
 \end{figure}     

Lastly, we study the influence of the bound state decay on the final densities of $\psi$ and $\chi$. The effect is illustrated in Fig.~\ref{fig:RPlot}
with again the same initial condition $\Delta Y^i_{\psi} = \Delta Y_B$.   
The decay eliminates the bound state population and stops BSD at an earlier time~(leads to higher $T_{\text{CU}}$) when compared to
situations of stable $\Rp$, since there are fewer bound states left over for dissociation.
That is to say, only BSF is active and causes more $\psi$ being converted into the bound states
that subsequently decay into $\chi$. Note that if the bound state decays only after BSF ceases to function, then
the final $\psi$ density will not be affected by the decay as displayed in Fig.~\ref{fig:exam_y}.

The decay width of $\Rp$ is partially controlled by the product of the couplings $\ka_\chi$ and $\ka_\psi$.
In Fig.~\ref{fig:RPlot}, the decay takes place during the catch-up period, and a larger decay width corresponds to fewer $\psi$ but more $\chi$ particles --
increasing the product $\ka_\chi \ka_\psi$ from $10^{-6}$ to $10^{-4}$ makes $Y^f_\psi$ more than ten times smaller.
Again, we focus on freeze-in scenarios where $\psi \psi~(\Rp)\leftrightarrow \chi \chi$ was not in thermal equilibrium at high $T$
but only {\it freezes in}  
during or after the catch-up period.
That imposes constraints on the parameter space as discussed in Appendix~\ref{sec:freeze-in_con}.

\subsection{Benchmark Scenarios} \label{sec:benchmark}
\begin{figure}
	\centering
		\includegraphics[width=0.48\textwidth]{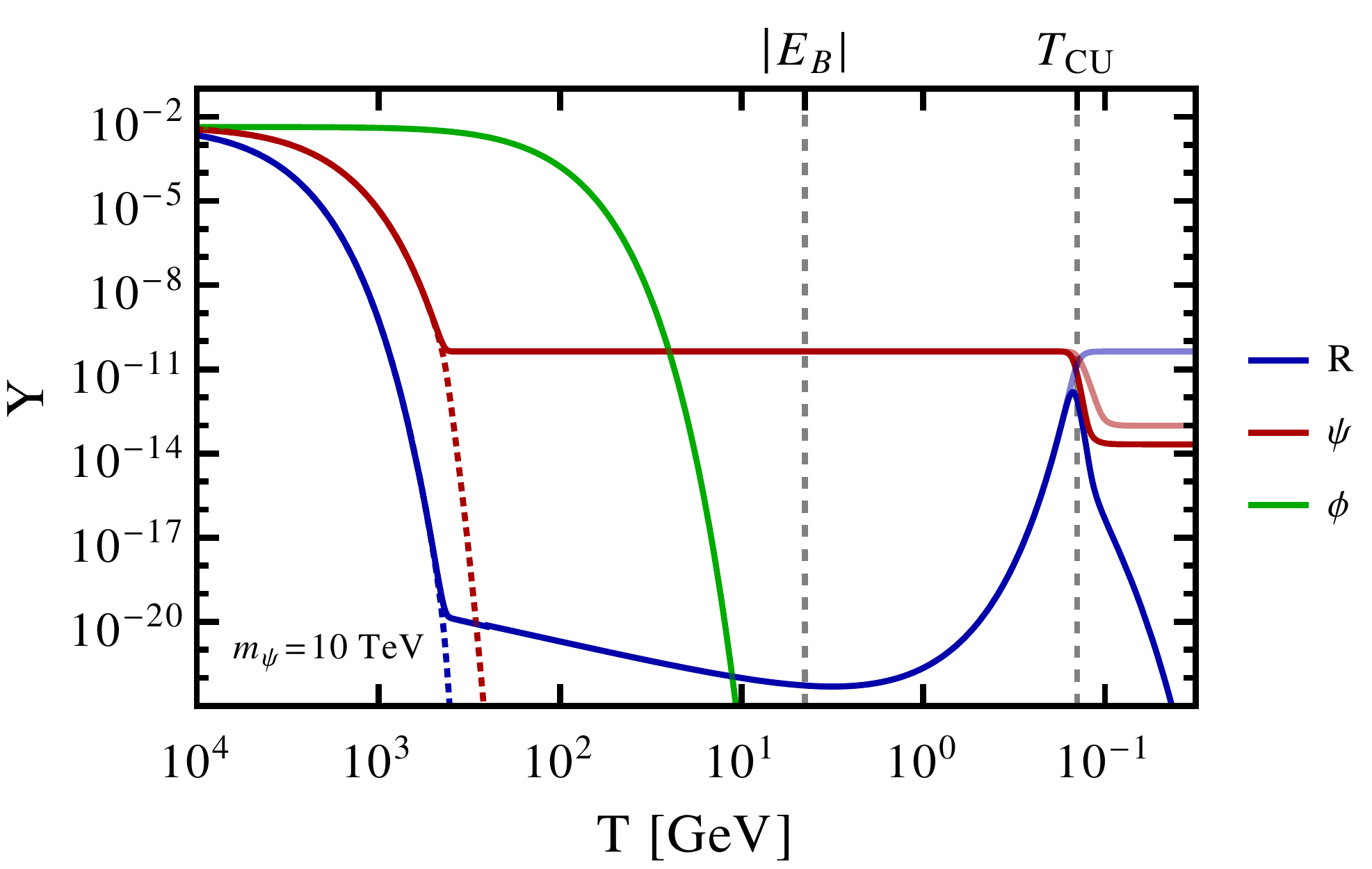}
		\includegraphics[width=0.48\textwidth]{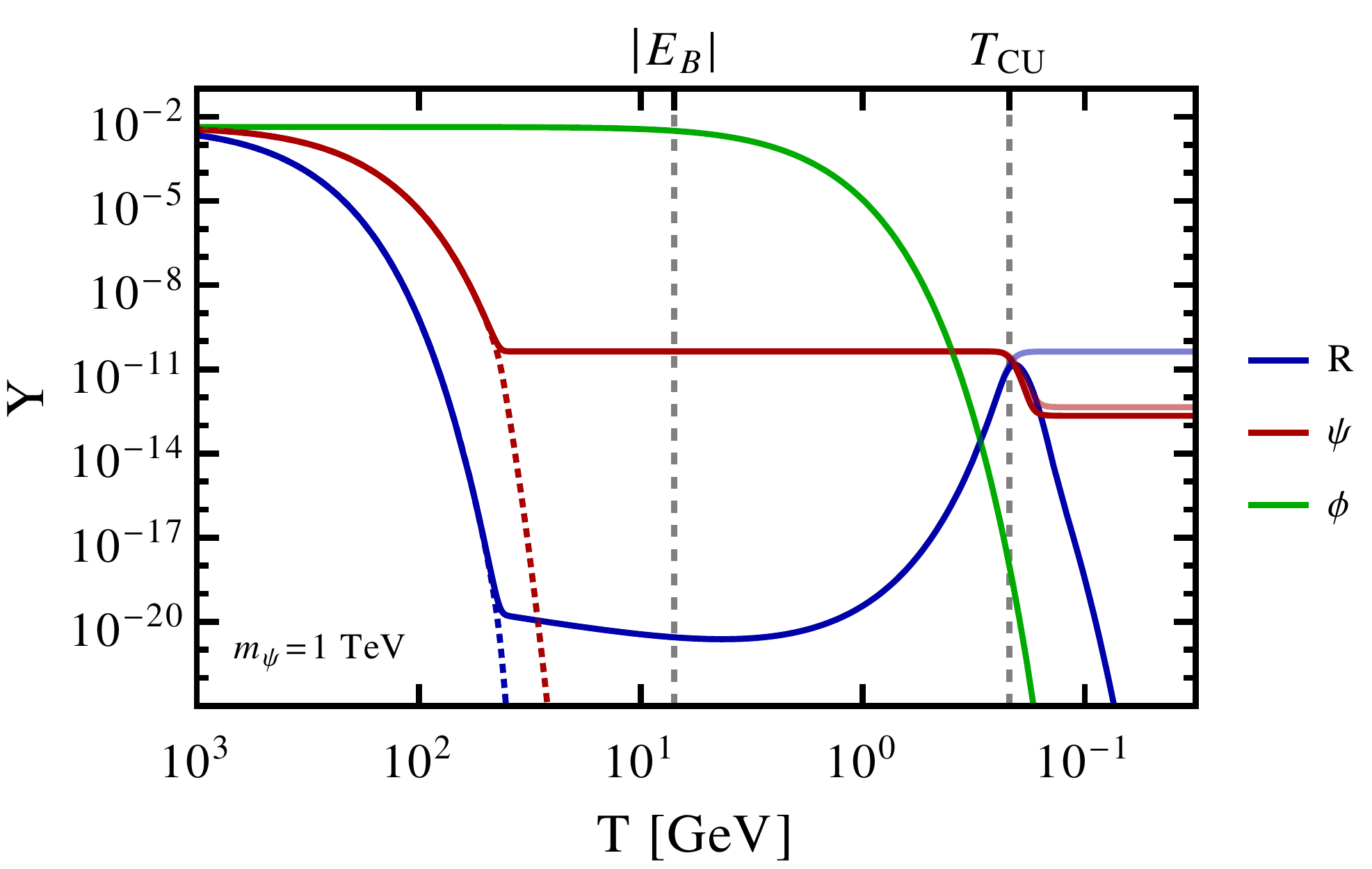}
		\includegraphics[width=0.48\textwidth]{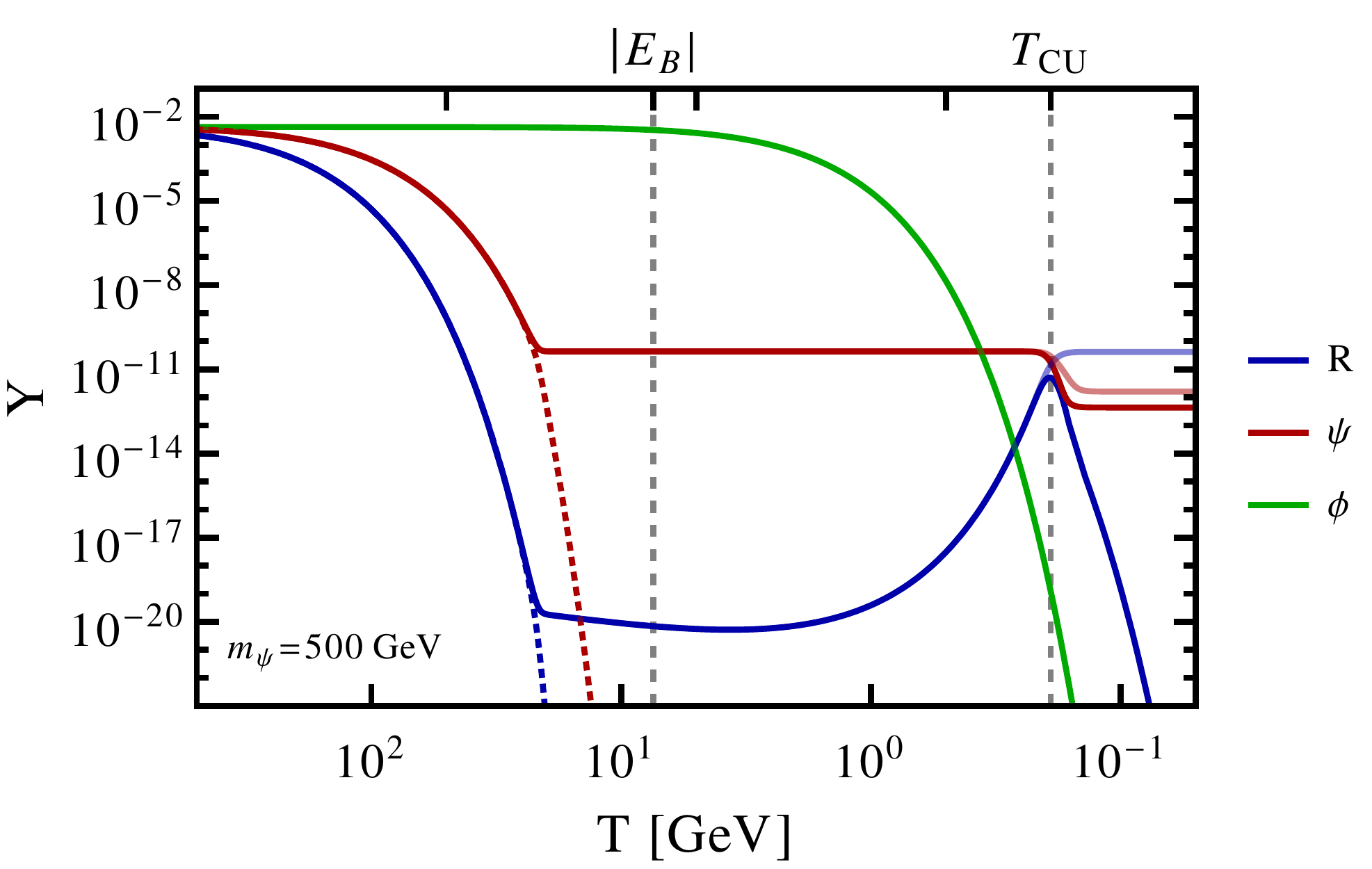}
	    \includegraphics[width=0.48\textwidth]{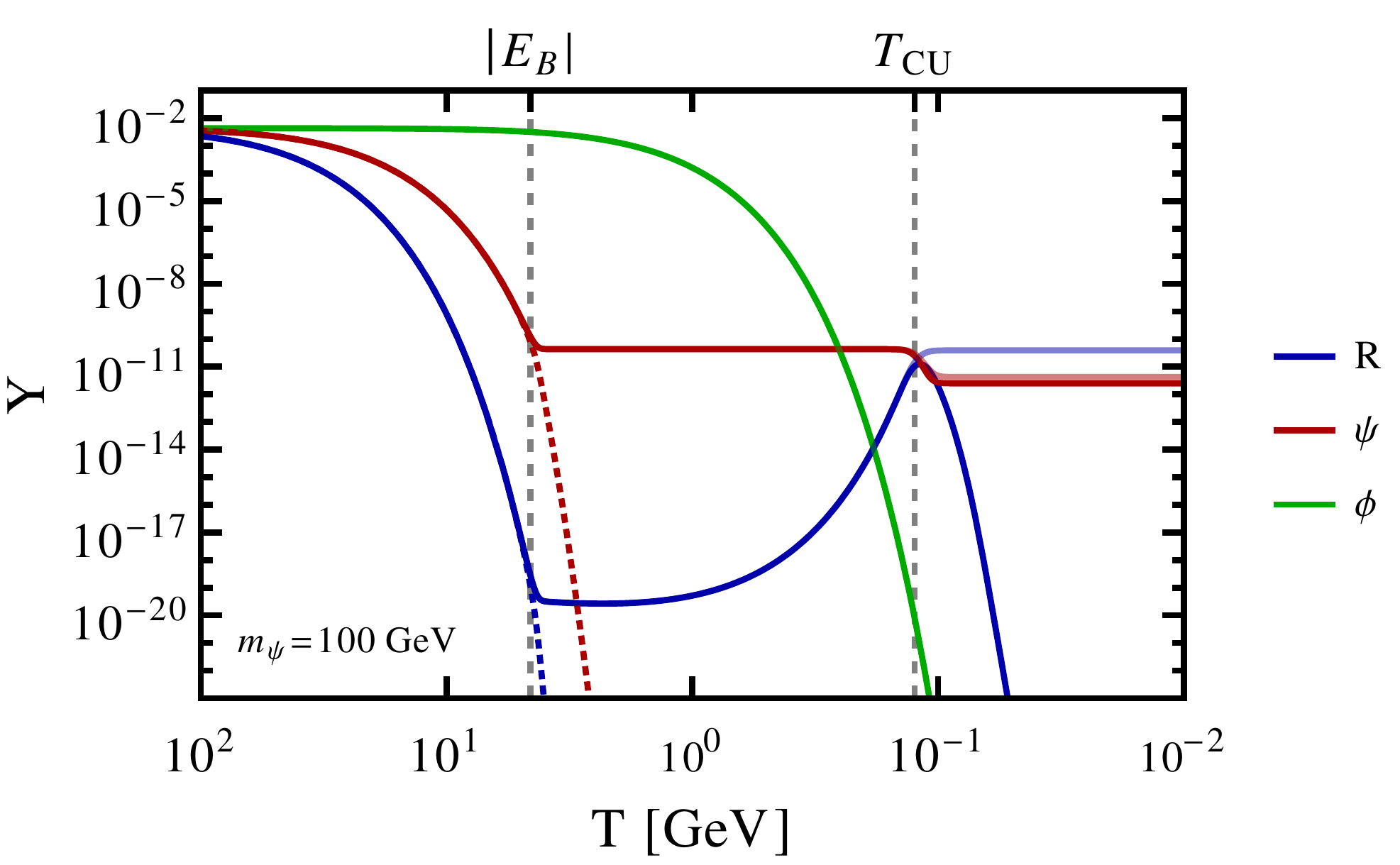}
	\caption{The results of the Boltzmann equations with $m_\psi=10 \, \mathrm{TeV}$~(upper left panel), $m_\psi=1 \, \mathrm{TeV}$~(upper right), $m_\psi=500 \, \mathrm{GeV}$~(bottom left) and  $m_\psi=100 \, \mathrm{GeV}$~(bottom right). These plots are similar to Fig.~\ref{fig:exam_y} but with $\Omega_\chi \approx \Omega_\psi$. We choose $\Delta Y_{\psi_i} = \Delta Y_{\chi_i} = - \Delta Y_B/2 $ as the initial condition at large $T$. The symmetric components of $\psi$ and $R$ are rapidly depleted via the annihilations into $\phi$ and the SM fermions. The relative distribution of the final asymmetric components of $\psi$ and $R$ is determined by when BSF and BSD decouple. Finally, the bound state decay shifts asymmetry into $\chi$. }
	\label{fig:results}
\end{figure}

To conclude, we present four benchmark sets of the parameters, listed in Table~\ref{tab:working}, which are capable of reproducing the observed $\Omega_{\text{DM}}$ with $\Omega_\chi \approx \Omega_\psi$.
\begin{table}[htp!]
	\centering
	\begin{tabular}{|c|c|c|c|c|c|c|}
		\hline
		$m_\chi$[GeV] & $m_\psi$[GeV] & $m_\phi$[GeV] & $m_{\phi'}$[GeV] & $y$ & $\kappa_\psi$ & $\kappa_\chi$    \\
		\hline
		$2.66$ & $10000$ & $5.7$ & $10000$ & $0.75$ & $3 \cdot 10^{-4}$  & $3 \cdot 10^{-4}$  \\
		$2.66$ & $1000$ & $9$ & $1000$ & $1.5$ & $1.2 \cdot 10^{-4}$  & $1.2 \cdot 10^{-4}$  \\
		$2.66$ & $500$ & $8.25$ & $500$ & $1.75$ & $1.5 \cdot 10^{-4}$  & $1.5 \cdot 10^{-4}$  \\
		$2.66$ & $100$ & $5.75$ & $100$ & $2.5$ & $7 \cdot 10^{-5}$  & $7 \cdot 10^{-5}$  \\
		\hline
	\end{tabular}
	\caption{Sets of parameters reproducing the observed relic density with the comparable energy densities between $\psi$ and $\chi$, assuming the initial condition $\Delta Y^i_{\chi} = \Delta Y^i_{\psi}=- \Delta Y_B/2$.}
	\label{tab:working}
\end{table}
The corresponding time evolution of the particle densities are shown in Fig.~\ref{fig:results}, similar to Fig.~\ref{fig:exam_y}
but with working values of the parameters.
The  mass of $\chi$ is fixed at $2.66$ GeV, while $m_\psi$ ranges from 100 GeV to 10 TeV.
It is assumed that the initial asymmetry created at $T \gg m_\psi$
is distributed as $\Delta Y^i_{\chi} = \Delta Y^i_{\psi}=- \Delta Y_B/2$. The rest of parameters are chosen
to fulfill $\Omega_\chi \approx \Omega_\psi$.

In order to have $\Rp$ decay before BBN, it is necessary for $T_\text{CU}$  to be above the BBN scale, i.e.
$T_\text{CU} \gtrsim$ MeV.
As the catch-up temperature $T_\text{CU}$ scales as $ \vert E_B \vert \, y^{1/5} \sim m_\psi \, y^{21/5}$,
for the cases with smaller $m_\psi$ a larger value of $y$ is requisite, as demonstrated in Table.~\ref{tab:working}.
However, due to the fact that the density $Y^f_\psi$ is extremely sensitive to $y$, as illustrated in Fig.~\ref{fig:yPlots}, 
a considerable value of $y$ will often make the final density of $\psi$ vanishingly small.
Therefore, as discussed in Section~\ref{sec:mediator}, a massive $\phi$~($m_\phi \gtrsim \vert E_B \vert$)
is involved to impede BSF, preventing  an utter depletion of $\psi$.
In fact, we have found for $m_\psi \lesssim \mathcal{O}$(TeV), massive $\phi$ is requisite to obtain $\Omega_\chi \sim \Omega_\psi$.

Lastly, the product $\kappa_\chi \kappa_\psi$, that determines the decay width of $\Rp$,
has to be sizable so that the decay will not interfere with BBN, but
it cannot be too large in order to retain the freeze-in decay.
In addition, the bound state decay leads to more $\psi$ being converted into $\chi$,
as seen by comparing the light~(without decay) and dark~(with decay) red lines in Fig.~\ref{fig:results}.

\section{Conclusions} \label{sec:con} 
As multi-component DM and asymmetric DM~(ADM) are interesting subjects on their own,
we here explore the possibility of combining these two ideas to have two-component ADM with very different masses:
the light component $\chi$  at the scale of GeV and the heavy component $\psi$ above the electroweak scale.
On the other hand, it is a common feature of existing ADM models that the baryon density~(asymmetry) is
closely correlated with that of DM, and quite often the amounts of asymmetry stored in the DM and SM sectors are of the same order, implying
the DM mass is of order $\mathcal{O}(\text{GeV})$ given $\Omega_{\text{DM}} = 5.4 \, \Omega_{B} $.
In the framework of two-component ADM, if an asymmetry created at an early time is
equally shared among $\chi$, $\psi$ and the SM baryons, 
the energy density of the heavy $\psi$ alone will certainly exceed  
the observed relic abundance.

A simple solution proposed in this work is to involve a Yukawa-type long-range interaction,
mediated by a scalar $\phi$, in the sector of $\psi$.
We always assume that the underlying mechanism of the asymmetry generation creates more $\chi$ and $\psi$ than $\bar{\chi}$ and $\bar{\psi}$, but our conclusions do not depend on this assumption.
Three types of bound states will form: $\Rp$, $\Rpb$, and $\Rm$, where the subscript denotes the bound
state constituents. The densities of the bound states and free $\psi$ are determined by the interplay between  bound state formation and dissociation~(BSF and BSD) $i + j \leftrightarrow R_{ij} + \phi$ for $(i,j) = \psi$ and/or $\bar{\psi}$. 
The presence of the bound states can facilitate removing the symmetric component of $\psi$, preserve the
asymmetric component and finally convert most of the asymmetry into $\chi$ via the late decay of $\Rp \to \chi \chi$.

To be more concrete, when the temperature becomes smaller than the mass of $\psi$, most of the symmetric component will be depleted and only the asymmetric component, consisting of $\psi$ and $\Rp$, remains with very different number
densities $n_\psi \gg n_{\Rp}$ due to the Boltzmann suppression.
As the temperature further drops below the binding energy of the bound state, $\phi$ no longer has sufficient
energy to break the bound states. Thus BSF is kinematically favored over BSD, making the density $n_{\Rp}$ catch up with $n_\psi$. As BSF proceeds, more and more $\psi$ particles have been converted and the process eventually stops because the interaction rate is proportional to $n^2_\psi$. In the mean time, the bound state starts to decay into $\chi$.
As a result, the final density of $\psi$ can be much smaller than that of $\chi$ while their energy
densities are comparable, i.e., two-component ADM.

However, the late decay of the bound states creates a population of energetic $\chi$ and injects entropy into the thermal bath
at $T \lesssim$ MeV, that will disturb BBN. To circumvent the issue, one can increase the Yukawa coupling, responsible for the long-range interaction, and involve a massive mediator $\phi$.
A larger coupling implies a larger binding energy that makes the bound state catch-up occur earlier.
In this case, the following decay of the bound state can take place before BBN. On the other hand,
a larger value of $y$ also implies a longer BSF epoch and a further depletion on the final density
of $\psi$~(analogous to thermal DM: a larger coupling between DM and SM particles implies a smaller relic density), resulting in
$\Omega_\psi \ll \Omega_\chi$ and thwarting the attempt on two-component ADM.
The undesirable effect of large $y$, nonetheless, can be undone by making $\phi$ massive.  
With the mass of $\phi$ being larger than the binding energy, it costs $\psi$ energy to form the bound states
due to $2\, m_\psi < m_{R} + m_\phi$ and therefore BSF will stop earlier, leaving a sizable population of $\psi$.

To conclude, we provide an interesting mechanism to achieve two-component ADM with very different masses but comparable energy densities by involving the long-range interaction and late decay in the heavy component sector.
In the follow-up, we will investigate phenomenological implications of this scenario, including boosted DM and DM searches in direct detection. 

\acknowledgments

We would like to thank Joachim Brod for precious contributions at the early stage of this work, and Martin Gorbahn for very helpful but destructive
comments on our previous ill-fated model. We are grateful for helpful comments on the draft from Yue-Lin Sming Tsai and Julia Harz.
WCH was supported by the Independent Research Fund Denmark, grant number 
DFF 6108-00623. The CP3-Origins centre is partially funded by the Danish National Research Foundation, 
grant number DNRF90.
MB would like to thank Heinrich P\"as for providing the possibility to work on this project and his constant support throughout the work.
 
\appendix

\section{Relevant reduced cross-sections and decay widths}
\label{app:reduced-X}
Here, we collect all relevant reduced cross-sections $\hat{\sigma}$, required for computing
the corresponding thermal rate $\gamma^{\text{eq}}$ used in the Boltzmann equations.
We here only consider CP-conserving tree-level processes, namely
$\gamma^\text{eq} \left( i \rightarrow f \right) = \gamma^\text{eq} \left( f \rightarrow i \right)$, whereas one in general has
$\gamma^\text{eq} \left( i \rightarrow f \right) = \gamma^\text{eq} \left( \bar{f} \rightarrow \bar{i} \right)$ according to
CPT invariance.
\begin{itemize}
	
	

	
	\item $\Rm \, \phi \to \psi \bar{\psi}$ \\
	\begin{align}
		\hat{\sig} \lee \Rm \phi \to \psi \bar{\psi} \rii = 2 s \lambda \left( 1 , \frac{m_{R}^2}{s} , \frac{m_{\phi}^2}{s} \right) \frac{2^4 y^{12} m_{\psi}^{\frac{5}{2}} m_R^5}{\left( s - m_R^2 - m_\phi^2 \right)^5} \sqrt{\frac{s + m_R \left( m_R - 4 m_{\psi} \right) - m_\phi^2}{2 m_R}} F \left( v \right) ,
	\end{align}
	with 
	\begin{align}
		F \left( v \right) = \frac{ v \exp \left(4 v \arctan \left[ v^{-1} \right] \right)}{\left( 1 - \exp \left( - 2 \pi v \right) \right) \left( 1 + v^2 \right)^2} \nn
	\end{align}
	and 
	\begin{align}
		v = \frac{y^2}{8 \pi} \sqrt{\frac{2 m_R m_{\psi}}{s + m_R^2 - 4 m_R m_{\psi} - m_\phi^2}} \, . \nn
	\end{align}
	Moreover, as the Yukawa interaction is always attractive among particles and antiparticles, one has
	\begin{align}
		\hat{\sig} \lee R_{\bar{\psi} \bar{\psi}} \phi \to \bar{\psi} \bar{\psi} \rii = \hat{\sig} \lee R_{\psi \psi} \phi \to \psi \psi \rii = 2 \hat{\sig} \lee \Rm \phi \to \bar{\psi} \psi  \rii  \, .
		\label{eq: sighat_BSF}
	\end{align}
	The factor of 2 for $\Rm$ can be understood as follows. 
The symmetry factor for $\Rp$ is $\frac{1}{2} \left( 2 \times 2 \right)^2$, where $1/2$ comes from the identical
outgoing particles~(phase-space integral reduced by $1/2$) and $\left( 2 \times 2 \right)$ is owing to the different ways for the Yukawa interaction to annihilate the initial state and create the
final state: $<\psi \psi | \phi \bar{\psi} \psi | \phi \psi \psi>$.
By contrast, $\Rm$ only has a symmetry factor of $\left( 2 \right)^2$ coming from two ways of annihilating and creating
the initial state and final state: $<\bar{\psi} \psi  | \phi \bar{\psi} \psi | \phi \bar{\psi} \psi >$.
As a consequence, there is a relative factor of 2 between the cases of $\Rp$ and $\Rm$.
	
	\item $\Rm \to \phi \phi$ \\
	
	The decay width of $\Rm$ at rest can be obtained by adapting the results from Eq.~(5.57) of Ref.~\cite{Peskin:1995ev}, where the bound state decay is induced by the massless photon. With simply replacing $\alpha_{em}^2$ with $\frac{y^2}{4 \pi}$, we have: 
	\begin{align}
		\Ga_{\Rm}= \frac{ 4 y^4}{ 79 \pi} \frac{| \Psi_{100}(0) |^2}{m_R^2}  \; ,
		\label{eq:Ga_rm}
	\end{align}
	where  $m_{\Rm} = m_\psi \lee 2 - \frac{y^4}{64 \pi^2} \rii$ 
	and $\Psi_{100}(0)$ is the ground state wave function of $\Rm$ at $r=0$.
	In the limit of $m_\phi \sim 0 \ll m_\psi$, the wave function reads
	\begin{align}
		\Psi_{100}(0) = \frac{y^3 \, m^{3/2}_\psi } { 16 \sqrt{2} \, \pi^2 } \; . \label{eq:ZeroWave}
	\end{align}
	
	\item $\Rp \to \chi \chi$ and $\Rpb \to \bar{\chi} \bar{\chi}$  \\
	The bound states $\Rp$ and $\Rpb$ have the same decay width.
	In the limit of $m_{\psi} \gg m_{\chi}$, it reads
	\begin{align}
		\Ga_{\Rp} = \Ga_{\Rpb} = \frac{ | \kappa_\chi \kappa_\psi |^2 | \Psi_{100}(0) |^2 m_{R}^2 }{12 \left( m^2_{\phi'} -  m^2_{R}\right)^2 \pi} , \label{eq: Ga_Rp}
	\end{align}
	where $\kappa_\psi$ and $\kappa_\chi$ are the couplings of $\psi$ and $\chi$ to the mediator $\phi'$
	in Eq.~\eqref{eq:rel_lan}, respectively, and $\Psi_{100}(0)$ is given by Eq.~\eqref{eq:ZeroWave}.
	
	\item $\psi \psi \leftrightarrow \phi \phi$  \\
	The reduced cross-section in the limit of  $m_\psi \gg m_\phi$ is given by
	\begin{align}
		\hat{\sig} \lee s \rii &= \frac{y^4}{4 \pi} \lee  \text{arctanh} \left[ \sqrt{1 - \frac{4 m_\psi^2}{s} } \right]   -  \sqrt{1 - \frac{4 m_\psi^2}{s}}  \vphantom{ \left[ \frac{\sqrt{s \left( s - 4 m_\psi^2 \right)}}{s} \right]} \rii \mathcal{S} \left( \zeta \right)  \, , 
	\end{align}
	with the Sommerfeld enhancement factor~\cite{ANDP:ANDP19314030302, Cirelli:2007xd}
	\begin{align}
		\mathcal{S} \left( \zeta \right) = \frac{2 \pi \zeta}{1 - \exp \left(- 2 \pi \zeta \right)} \,  , \nn
	\end{align}
	and
	\begin{align}
		\zeta = \frac{y^2}{4 \pi} \frac{1}{\sqrt{1 - \frac{4 m_\psi^2}{s}}} \, . \nn
	\end{align}

\end{itemize}


\section{Bound State Formation and Dissociation}
\label{app:Bound_F_D}
In this Section, we discuss the cross-section computation of bound state dissociation~(BSD) for the bound state $\Rp$
by closely following the formalism described in Ref.~\cite{Landau:1982ev}.
The bound state formation~(BSF) rate can be straightforwardly derived from that of BSD
via  $\gamma^\text{eq} \left( \psi \psi \to \Rp  \phi \right) = \gamma^\text{eq} \left( \Rp  \phi  \to \psi \psi \right) $, 
while Eq.~\eqref{eq: sighat_BSF} can be used to infer the BSD rates for the other bound states, $\Rm$ and $\Rpb$.

To compute the amplitude of BSD, one needs to know the wave-function overlap between the initial and final states. Therefore, the bound state wave function in the presence of a Yukawa potential is required.
In general, it does not have an analytic expression~(see, e.g., Refs~\cite{Hamzavi_2012,PhysRevA.1.1577}). 
To simplify the calculation, in the following we will focus on regions of the parameter space where the Yukawa potential can be
well approximated by a Coulomb potential, of which the wave function is well-known.

\subsection{Non relativistic Case}
We start with the case of non-relativistic $\psi$. The Yukawa potential is given by
\begin{align}
	V \left( r \right) = -\frac{y^2}{4 \pi} \frac{\exp \left( - m_\phi r
		\right)}{ r} = -\frac{y^2}{4 \pi} \frac{1}{r} \bigg[ 1 - m_\phi r +
	{\mathcal O}\big((m_\phi r)^2\big) \bigg]\,,
\end{align}  
where $y$ is the Yukawa coupling and $m_{\phi}$ is the mass of the
scalar mediator $\phi$. In case the mediator mass is much smaller than the inverse of the
Bohr radius $a_0~(= 8\pi / (y^2  m_\psi))$,
the Yukawa potential will be dominated by the leading term since $m_\phi r \sim m_\phi a_0 \ll 1$,
leading to a Coulomb potential.  
Under this approximation, we
can solve the Schr\"odinger equation for the Coulomb potential and
obtain the ground state wave function
\begin{equation}
	\Psi_i \left( r \right) = \frac{m_\psi^{3/2} y^3}{16 \sqrt{2} \pi^2}
	\exp \left( - \frac{m_\psi y^2}{8 \pi} r \right) \, , \label{eq:GroundState}
\end{equation}
where the subscript $i$ refers to the initial state, as well as the binding energy 
\begin{equation}
	E_B = - \frac{m_\psi y^4}{64 \pi^2} + \frac{m_\phi y^2}{4 \pi} \,.
\end{equation}
The differential cross-section for the process of $\Rp + \phi
\rightarrow \psi + \psi$ is given by
\begin{equation}
	\frac{d \sigma}{d \Omega} = \frac{|V_{fi}|^2}{ \left( 2 \pi \right)^2
	} \mu_\psi |{\bf p}|\,, \label{eq:dSigma}
\end{equation}
where ${\bf p} \equiv \mu_\psi ({\bf p}_{\psi,1}/m_\psi - {\bf
	p}_{\psi,2}/m_\psi)$ 
	is the relative momentum between the two $\psi$
particles. 
Conservation of energy requires $|{\bf p}| = \sqrt{2\mu_\psi
	(E_B + E_\phi)}$ and the matrix element $V_{fi}$ is defined as:
\begin{align}
	V_{fi} = y \sqrt{\frac{2 \pi }{E_{\phi}}} \int \Psi_{i}^* \exp \left( i k r \right) \Psi_{f} \equiv 
	y \sqrt{\frac{2 \pi }{E_{\phi}}} M_{fi} \, , \label{eq:Vfi}
\end{align}
where $k$ is the momentum of the $\phi$ particle.
In contrast to the matrix element presented in Chapter 56 of
\cite{Landau:1982ev}, a Yukawa-type interaction, $ \mathcal{L} \supset y \phi \bar{\psi} \psi$, is considered here instead of
the Coulomb interaction mediated by the photon, $\mathcal{L} \supset e \bar{\psi}
\gamma_{\mu} \psi A^{\mu}$.
Since the computation assumes the unbound $\psi$ to be non-relativistic it is sufficient to use the solution of the Schr\"odinger equation with a positive energy eigenvalue for describing the unbound final state $\Psi_f$: 
\begin{align}
	\Psi_{f}&= \frac{m_{\psi} y^2}{4 \sqrt{2 \pi} |{\bf p}|} \frac{\exp \left( -i |{\bf p}| r \right)}{\sqrt{ v \left[1 - \exp \left( 2 \pi v \right) \right]}} {}_1F_1 \left( 1 + iv, 2 , 2 i |{\bf p}| r \right).
\label{eq:freeState}	 
\end{align}
Here, we have $v = y^2 m / (8 \pi |{\bf p}| )$ and only the $l=0$ component is included due to the angular momentum conservation.
Furthermore, we assume $\exp \left( i k r \right) \approx 1$, which is a good approximation
as long as the assumption of the Coulomb potential is valid, i.e., $m_\phi \ll y^2 m_\psi / ( 8\pi )$.  
From Eqs.~\eqref{eq:GroundState} and \eqref{eq:freeState}, the integral in Eq.~\eqref{eq:Vfi} becomes
\begin{align}
	V_{fi} = - \frac{y^6 \sqrt{m_{\psi}}}{2 \sqrt{2} \pi E_{\phi}^{\frac{5}{2}}} \sqrt{\frac{v}{1 - \exp \left( - 2 \pi v \right)}} \frac{\exp \left( 2 v \arctan \left[ v^{-1} \right] \right)}{1 + v^2} .  
\end{align}
Note that we have made a replacement of $r \rightarrow 2r$ in the wave functions $\psi_i$ and $\psi_f$ in the integral \eqref{eq:Vfi}, since $dr$ is defined as the position relative to the center of mass of the bound state, whereas the relative position is used before  in the bound state wave functions \eqref{eq:GroundState} and \eqref{eq:freeState}. Since we are dealing with a bound state consisting of two particles of equal mass, there is a factor of 2 difference between these two quantities. 

Finally, by integrating Eq.~\eqref{eq:dSigma} over the solid angle and employing the conservation of the kinetic energy, $E_{\phi} = \frac{|{\bf p}|^2}{2 \mu} + \frac{m y^2}{64 \pi^2} = \frac{|{\bf p}|^2}{m_{\psi}} \left( 1 + v^2 \right)$,
the cross-section for the non-relativistic BSD is obtained:
\begin{align}
\sigma = \frac{y^{12} m_{\psi}^{\frac{5}{2}} \sqrt{E_B + E_{\phi}}}{2 \pi^3 E_{\phi}^5} \frac{ v \exp \left(4 v \arctan \left[ v^{-1} \right] \right)}{\left( 1 - \exp \left( - 2 \pi v \right) \right) \left( 1 + v^2 \right)^2}.
\label{eq:BSD_sig}
\end{align}
For BSD of $\Rm$, one has to include an additional factor of $1/2$ to Eq.~\eqref{eq:BSD_sig} according to
Eq.~\eqref{eq: sighat_BSF}.
Moreover, it is more convenient to rewrite the result in terms of  the center-of-mass energy $s$
in order to apply Eq.~\eqref{eq:ga_eq}. In the rest frame of the bound state, the center of mass energy is $s = m_R^2 + 2 m_R E_{\phi}+ m_\phi^2$.
Thus, the cross-section becomes:
\begin{align}
	\sigma \left( s \right) =  \frac{2^4 y^{12} m_{\psi}^{\frac{5}{2}} m_R^5}{\left( s - m_R^2 - m_\phi^2 \right)^5} \sqrt{\frac{s + m_R \left( m_R - 4 m_{\psi} \right) - m_\phi^2}{2 m_R}} \frac{ v \exp \left(4 v \arctan \left[ v^{-1} \right] \right)}{\left( 1 - \exp \left( - 2 \pi v \right) \right) \left( 1+ v^2 \right)^2} ,
\end{align}
with 
\begin{align}
v = \frac{y^2  m_{\psi} }{8 \pi |{\bf p}|} = \frac{y^2}{8 \pi} \sqrt{\frac{2 m_{\psi} m_R}{s + m_R \left( m_R - 4 m_{\psi} \right) - m_\phi^2}}
\, . \nn
\end{align}

\subsection{Relativistic Case}
For the case of relativistic $\psi$, we follow Chapter 57 of Ref.~\cite{Landau:1982ev}, where results of
the hydrogen atom have to be adapted for the Yukawa coupling as above.
To calculate the matrix element $M_{fi}$, defined in Eq.~\eqref{eq:Vfi}, the initial and final state wave functions are required.
The unbound, outgoing $\psi$ is assumed to be highly relativistic. Therefore, the wave function is taken to be a plane wave:  
\begin{align}
	\psi_{f} = \sqrt{\frac{1}{2 E_{\psi}}} u_{f} \exp \left( i  p  r \right) .  
\end{align}
Since the initial state is also relativistic now, the first-order relativistic correction should be included: 
\begin{align}
	\psi_{i} = \left( 1 - \frac{i}{2 \mu_{\psi}} \gamma^0 \vec{\gamma} \vec{\bigtriangledown} \right) \frac{u_{i}}{\sqrt{2 \mu_{\psi}}} \psi_{nr} \, ,
\end{align}
where the wave function is derived in Chapter 39 of Ref.~\cite{Landau:1982ev} and $\psi_{nr}$ is simply
the ground state wave function in Eq.~\eqref{eq:GroundState}.
Substituting these equations into \eqref{eq:Vfi} yields 
\begin{align}
	M_{fi} = \frac{1}{2 \sqrt{\mu E_{ \psi}}} \int d^3 x \, \bar{u}_f \left( \gamma^0 - \frac{i}{2 \mu_{\psi}} \vec{\gamma} \vec{\bigtriangledown} \right) u_i \psi_{nr} e^{-i \left( \bf p - \bf k \right) \bf r} \, ,
\end{align}
that results in
\begin{align}
	|M_{fi}|^2 = \frac{y^{10} m_{\psi}^4}{256 \pi^4 E_{\psi} \left( \bf p - \bf k \right)^4} \bar{u}_f A u_i \left( \bar{u}_f A u_i \right)^{\dagger}  \,  ,
\end{align}
with
\begin{align}
	A = \frac{\gamma^0}{\left( \bf p - \bf k \right)^2}+ {\boldsymbol \gamma} \frac{ \bf k -  \bf p}{2  \mu_{\psi} \left( \bf k - \bf p \right)^2} \, . 
\end{align}
Here $\bf{k}$ corresponds to the momentum of $\phi$ and $\bf{p}$ stands for the momentum of the unbound $\psi$ in the rest frame of the bound state before the collision. 
After summing over the final spins and averaging over the initial ones, we obtain
\begin{align}
	\frac{d \sigma}{d \Omega} = \frac{ y^{12} m_{\psi}^5 |\bf p|}{256  \pi^5 E_{\phi} \left( { \bf p} - {\bf k} \right)^6} \left( \frac{E_{\psi} + m_{\psi}}{\left( {\bf p} - {\bf k} \right)^2} + \frac{E_{\psi} }{m_{\psi}^2} - \frac{{\bf p}^2 - {\bf k}^2}{m_{\psi} \left( {\bf p} - {\bf k} \right)^2} \right) \, . \label{eq:BSDrel}
\end{align}
In contrast to the case of the hydrogen atom, neither of the two particles forming the bound state can be treated at rest in this system. We have to first calculate $\bf{k'}$ and $\bf{p'}$ in the center-of-mass system of the collision, and then perform a Lorentz boost back into the rest frame of the bound state afterwards. The procedures are lengthy but straightforward,
and will not be shown here. 
Additionally, the integral over the solid angle $d\, \Omega$ in eq. \eqref{eq:BSDrel} can only be computed numerically.
The resulting cross-section is a function of the center-of-mass
energy~($s = m_R^2 + 2 m_R E_{\phi} + m_\phi^2$), the mass of $\psi$ and the Yukawa coupling. 

\section{Freeze-in Constrain on the interaction rate of $\chi \chi \leftrightarrow \psi \psi$} \label{sec:freeze-in_con}

As emphasized in Section~\ref{sec:introduction},
we focus on the scenario of freeze-in decay,
where the bound state starts to decay~(via $\psi \psi \to \chi \chi$) only during or after
the density of the bound state caught up with that of the free $\psi$. 
On the other hand, the process $\chi \chi \leftrightarrow \psi \psi$ also transfers asymmetry from $\psi$ to
$\chi$ if it is efficient~($\Gamma_{\psi \psi \leftrightarrow \chi \chi} > H$) for temperatures below the mass of $\psi$. 
As our main goal in this work is to demonstrate that BSF and the freeze-in decay
can preserve and convert asymmetry
to attain two-component ADM, we have to make sure the aforementioned process is not in equilibrium    
before the freeze-in decay. Here, we discuss how to satisfy the out-of-equilibrium constraint. 

The reduced cross-section of the process $\chi \chi \leftrightarrow \psi \psi$ is given by:
\begin{align}
	\hat{\sig} = \frac{s^2 \sqrt{1 - \frac{4 m_\psi^2}{s}}}{8 \pi \left[ \left( s - m_{\phip}^2 \right)^2 + m_{\phip}^2 \Gamma_{\phip}^2 \right]} \, ,
\end{align}
and the thermal rate can be found from
\begin{align}
	\ga^\text{eq} \lee a_1 a_2 \toto f_1 f_2 \rii = \frac{T}{64 \pi^4} \int^{\infty}_{s_{\text{min}}} ds \, \sqrt{s} \, 
	\hat{\sigma}(s) \, K_1\lee \frac{\sqrt{s}}{T} \rii \, . \nn
\end{align}
In case $m_{\phip}>2 m_\psi$ and a small decay width, the narrow width approximation can be applied to evaluate the integral and we have
\begin{align}
	\gamma^\text{eq} \approx \frac{\kappa_\chi^2 \kappa_\psi^2 m_{\phip}^3 T \sqrt{m_{\phip}^2 - 4 m_\psi^2}}{2^9 \pi^5 \Gamma_{\phip}} K_1 \left( \frac{m_{\phip}}{T} \right) \, .
\end{align} 
To determine whether or not the process is in equilibrium, we need to compute the ratio of
its interaction rate to the Hubble expansion rate. It reads 
\begin{align}
	\frac{\Gamma_{\psi \psi \leftrightarrow \chi \chi}}{H} \approx 3.6 \cdot 10^{-6} \frac{\kappa_\chi^2 \kappa_\psi^2 M_\text{Pl} m_{\phip}^3 \sqrt{m_{\phip}^2 - 4 m_\psi^2} \, K_1 \left( \frac{m_{\phip}}{T} \right)}{m_\psi^2 T^3 \Gamma_{\phip}} \, , 
	\label{eq:estim}
\end{align}
which is required to be less than one for the out-of-equilibrium condition at $T \lesssim m_\psi$.
If the mediator $\phi'$ only couples to $\chi$ and $\psi$, then the decay width is simply
\begin{align}
\Gamma_{\phip} = \frac{m_{\phip}}{8 \pi} ( \kappa_\chi^2 + \kappa_\psi^2 ) \, .
\end{align}
In this case, although the interaction rate is suppressed by a small factor of $\ka^2_\chi \ka^2_\psi$, 
it is at the same time enhanced by the small decay width proportional to $\ka^2_{\chi,\psi}$.
All in all, the interaction rate is only suppressed by the square of the couplings $\ka_{\chi,\psi}$.
Note that the bound state decay width is also proportional $\ka_{\chi,\psi}^2$ and
demanding the decay to happen before BBN imposes a lower bound on values of the couplings.
In this case, the suppression of $\kappa^2$ is not strong enough to
keep the process $\psi \psi \leftrightarrow \chi \chi$ out of equilibrium at  $T \lesssim m_\psi$.  
One solution is to involve additional decay channels for $\phip$ that increase the total decay width and hence
weaken the resonance enhancement. In this way, the interaction rate is suppressed by $\ka^2_\chi \ka^2_\psi$,
and with reasonably small values of the couplings~(without perturbing BBN) the out-of-equilibrium condition can be fulfilled. 
Alternatively, one can simply make $m_{\phip} < 2 \, m_\psi$ to prevent the resonance enhancement.
For the benchmark points listed  in Table.~\ref{tab:working},
the second solution is utilized by taking $m_{\phip} = m_\psi$ and we
have numerically confirmed that all the points satisfy the decoupling constraint.

\bibliography{two_ADM}
\bibliographystyle{hunsrt}

\end{document}